\documentclass[aps,pra,preprint,showpacs,superscriptaddress,floatfix]{revtex4-1}
\usepackage{graphicx}
\usepackage{amsmath}
\usepackage{amsfonts}
\usepackage{amssymb}
\usepackage{booktabs}
\usepackage[utf8]{inputenc}
\usepackage{subcaption}

\usepackage{bm}
\begin{document}

\title{First-order symmetry-adapted perturbation theory for multiplet splittings}

\author{Konrad Patkowski}
\affiliation{Department of Chemistry and Biochemistry, Auburn University,
Auburn, AL 36849, United States}
\author{Piotr S. \.Zuchowski}
\affiliation{Institute of Physics, Faculty of Physics, Astronomy and Informatics, 
Nicolaus Copernicus University, Grudziadzka 5, Torun 87-100, Poland}
\author{Daniel G. A. Smith}
\altaffiliation{Current address: The Molecular Sciences Software Institute,
1880 Pratt Drive, Suite 1100,
Blacksburg, VA 24060, United States}
\affiliation{Department of Chemistry and Biochemistry, Auburn University,
Auburn, AL 36849, United States}

\date{\today}

\begin{abstract}
We present a symmetry-adapted perturbation theory (SAPT) for the interaction of two high-spin open-shell molecules (described by their
restricted open-shell Hartree-Fock determinants) resulting in low-spin states of the complex. The previously available SAPT formalisms,
except for some system-specific studies for few-electron complexes, were restricted to the high-spin state of the interacting system. Thus,
the new approach provides, for the first time, a SAPT-based estimate of the splittings between different spin states of the complex. We have
derived and implemented the lowest-order SAPT term responsible for these splittings, that is, the first-order exchange energy. We show that
within the so-called $S^2$ approximation commonly used in SAPT (neglecting 
effects that vanish as fourth or higher powers of intermolecular overlap integrals), the first-order exchange energies for
all multiplets are linear combinations of two matrix elements: a {\em diagonal exchange term} that determines the spin-averaged effect and a
{\em spin-flip term} responsible for the splittings between the states. The numerical factors in this linear combination are determined
solely by the Clebsch-Gordan coefficients: accordingly, the $S^2$ approximation implies a Heisenberg Hamiltonian picture with a single coupling
strength parameter determining all the splittings. The new approach is cast into both molecular-orbital and atomic-orbital expressions: the
latter enable an efficient density-fitted implementation. We test the newly developed formalism on several open-shell complexes ranging from
diatomic systems (Li$\cdots$H, Mn$\cdots$Mn, \ldots) to the phenalenyl dimer.
\end{abstract}

\maketitle

\section{Introduction}

The interactions of open-shell atoms, molecules, and ions are of great significance to many areas of molecular physics and chemistry. 
As a matter of fact, all chemical reactions at some stage have to experience bond breaking and rearrangement and to describe such processes one has to deal with interacting fragments of reactants or products. 
Many electronically excited states of molecules have open electronic shells, so in studies of the interactions of such molecules with background gas or solvent similar problems arise. 
Interacting radicals are of key importance in atmosphere, where they undergo chain reactions, form metastable states and stable complexes \cite{Aloisio:00,Galano:10}. 
Collisions of radicals (such as OH, CH$_2$, CN, and many more) with hydrogen (H and H$_2$)  are a subject of study for astrochemistry \cite{Roueff:13} as they provide information about the conditions which molecules experience in the interstellar clouds. 
Stabilized organic radicals are relevant in organic and materials chemistry~\cite{Ratera:12}, for example, as building blocks  of molecular magnets. 
One should also mention the high relevance of interactions of radicals in cold chemistry: open-shell molecules can be manipulated with magnetic fields to control their collisions at low temperatures, giving unique information on the interaction potential \cite{Kirste:12,Lavert-Ofir:14,Chefdeville:13}.

In modelling the dynamics of weakly interacting systems, accurate potential energy surfaces are crucial. In case of open-shell systems, despite large progress in recent years, electronic structure calculations are still far from routine or robust.  
Although  the  coupled cluster (CC) method, most widely used in studying molecular interactions (in particular in its variant with single, double, and perturbative triple excitations, CCSD(T)),   has been extended to the high-spin open-shell case a while ago by Knowles et al.~\cite{Knowles:93} and Watts et al.~\cite{Watts:93}, there are many examples where the coupled cluster method fails and a multireference approach is necessary. 
In particular, one of the most important cases for which the single-reference CC method does not work is the dissociation of a low-spin system into high-spin subsystems: for example, a breakdown of a singlet, stable molecule into two doublet molecules. 
However, despite many attempts and  progress in theory since the introduction of the multireference coupled-cluster (MRCC) ansatz by Jeziorski and  Monkhorst~\cite{Jeziorski:81}, the proposed variations of MRCC are far from being black-box methods (see Refs.~\cite{Bartlett:07} and \cite{Jeziorski:10} for a review).  
At present, to account for both static and dynamical correlation effects in molecular interactions for systems exhibiting a non-unique electronic configuration, multireference configuration interaction~ \cite{Werner:88}~ (MRCI) or multireference perturbation theories \cite{Andersson:90,Finley:98,Angeli:02}  are often applied.  
In the case of the MRCI method, it is possible to reproduce the short-range part of the potential well when size-consistency corrections are applied. 
Such {\em a posteriori} corrections are not exact  and non-unique, and the remaining size-consistency error might be on the order of the interaction energy. 
In case of the perturbation methods, the same problem also arises in general; moreover, some of these methods suffer from the problem of intruder states \cite{Camacho:10}. 
One should also mention that all multireference methods rely on a proper selection of the active space, which is not necessarily straightforward, and that the complexity of calculations grows exponentially with the size of the active space. This is a particularly severe problem for interacting open-shell molecules as the active space required for the complex, typically the union of the monomers' active spaces, might be intractable. 

An interesting alternative for open-shell systems is the family of the so-called spin-flip  methods in which it is possible to reach arbitrary, multireference low-spin states of a given system starting from a single-reference high-spin state~\cite{Krylov:01}. 
The spin-flip equation of motion approach has been implemented within a variety of different theories, in particular, Hartree-Fock~\cite{Krylov:01}, density functional theory~\cite{Shao:03}, configuration interaction doubles~\cite{Krylov:01} and finally, using the  CC framework~\cite{Levchenko:04,Krylov:06}. The theory has been generalized to multiple spin flips \cite{Zimmerman:12}, however, the resulting calculations can be quite demanding, and it is often advantageous to describe the exchange splittings between all multiplets by a single exchange coupling constant $J_{AB}$ within the Heisenberg spin Hamiltonian. In order to determine $J_{AB}$, it is sufficient to perform a single spin-flip calculation starting from the high-spin, single-determinant reference state \cite{Mayhall:15}. 

Recent progress notwithstanding, the current arsenal  of methods which can be used for multireference open-shell complexes is limited, and it is desirable to explore alternative approaches which provide good insight and reliable interaction potentials.  
The closed-shell symmetry-adapted perturbation theory (SAPT)~\cite{Jeziorski:94,Hohenstein:12} has been widely recognized as a highly useful tool for calculations of interaction energies with an accuracy  comparable to CCSD(T), but also as a robust analysis tool which provides insightful physical interpretation of the nature of the interaction in terms of its electrostatic, induction, dispersion, and exchange components. 
For general single-reference open-shell systems,  \.Zuchowski et al. \cite{Zuchowski:08} and Hapka et al. \cite{Hapka:12}  introduced the symmetry-adapted perturbation theory based on, respectively, restricted and unrestricted  Slater determinants (with Hartree-Fock or Kohn-Sham orbitals). 
Present implementations of SAPT, however, are valid only for the single-reference high-spin case of the dimer in which the spin quantum number of the complex is equal to the sum of spin quantum numbers of monomers ($S = S_{\rm A} + S_{\rm B}$).

The interaction of two open-shell species with the total spin of the dimer smaller than the sum of spin quantum numbers of the monomers still poses a challenge.   
This is the case since low-spin configurations are typical examples of multireference systems. The knowledge of the so-called exchange splitting, which is defined as 
the difference between the highest and lowest possible spin state of a given dimer, is very important in interactions of small open-shell molecules (eg. in the O$_2$~$^3\Sigma^-_g$ dimer~\cite{Aquilanti:99,Bartolomei:08,Bartolomei:10})
as well as for interactions involving stable organic radicals such as phenalenyl~\cite{Cui:14}  or unsaturated metal sites within metal organic frameworks in the O$_2$ adsorption process \cite{Parkes:15}. 
The new approaches of calculating low-spin potential energy surfaces could be applied to model  fragments of interaction potentials in reaction dynamics, for instance near the  channels  which correspond to dissociation into two radicals, or near the  bond-breaking geometry of the reactive complex. 
Spin splitting also plays an important role in cold physics, since it is the term that strongly couples hyperfine states of alkali-metal atoms and drives Feshbach resonances \cite{Bartenstein:05,Chin:04}.   
It should be stressed that in the SAPT approach exchange splitting does indeed come exclusively from exchange corrections: the electrostatic, induction, and dispersion SAPT terms are identical for all asymptotically degenerate states of different spin multiplicity.

This paper is the first step toward the development of SAPT for low-spin dimer states. Here, we derive the first-order exchange energy in terms of spin-restricted  orbitals which preserve the correct value  of the squared-spin operator $\widehat{S}^2$. 
The theory introduced in this paper is beneficial for several reasons:  (i) as usually in SAPT, the interaction energy components are calculated {\em directly}, which contrasts with supermolecular calculations that always involve subtraction; (ii) there is  no need for the active space selection for the complex: in fact, no multireference calculation is necessary (similar to the spin-flip electronic structure approaches); (iii) it is possible to calculate the energy in the monomer-centered  basis set,  and the overall scaling of the method (once transformed molecular-orbital (MO) integrals are available) is just $o^4$ where $o$ is the number of occupied orbitals; 
and (iv) the low-spin exchange energy component, together with other SAPT corrections, can provide an insight into the nature of the interaction in a radical-radical system. 
The formalism presented here is valid for any spin-restricted Slater determinants. At this level, similarly to the high-spin open-shell theory \cite{Zuchowski:08}, intramonomer correlation effects are not included except for the possible use of Kohn-Sham orbitals \cite{Williams:01}. As we will show in Sec.~\ref{theory}, the
first-order SAPT term responsible for the multiplet splittings involves matrix elements between the
zeroth-order wavefunction and a function where one of the unpaired spins on one monomer has been lowered and
one of the unpaired spins on the other monomer has been raised, in other words, a function which was
obtained from the zeroth-order SAPT wavefunction via an intermolecular spin flip. Therefore, we will
refer to our new approach as spin-flip SAPT (SF-SAPT), borrowing the nomenclature from spin-flip
electronic structure theories \cite{Krylov:01}. However, while in the latter theories the $z$ component of the total
spin of the system (the $M_S$ quantum number) is changed upon the spin flip, in SF-SAPT only the $M_{S_A}$ and $M_{S_B}$ numbers for individual molecules
are changed: the overall $M_S$ value is conserved.

To date, there are only a few applications of perturbation theory to low-spin complexes. 
In 1984, Wormer and van der Avoird used the Heitler-London formula with appropriately spin-adapted linear combinations of Slater determinants (in the specific case of 4 active electrons) to predict the splittings between different multiplets of the O$_2\cdots$O$_2$ system \cite{Wormer:84}. 
Extensive studies of SAPT performance up to very high order of perturbation theory exist for few-electron systems in which monomers  can be represented by exact (in a given Gaussian basis) wavefunctions.  
\'Cwiok et al. \cite{Cwiok:92} studied the singlet and triplet states of the  H$\cdots$H interaction, while Patkowski et al. reported the convergence of various SAPT variants for the lowest singlet and triplet states of the Li$\cdots$H complex \cite{Patkowski:01, Patkowski:02}.

In the past few years, the exchange splitting of interaction energy has been studied for the H$_2^+$ system by Gniewek and Jeziorski~\cite{Gniewek:14,Gniewek:15,Gniewek:16}. Although this is a single-electron system, its two lowest  states are asymptotically degenerate and correspond to symmetric ($g$) and antisymmetric  ($u$) combinations of atomic orbitals. 
The difference between $u$ and $g$ states results from the resonance tunneling of the electron
between the nuclei, in a similar way 
as the difference between triplet and singlet states in the interaction of two doublets. 
Gniewek and Jeziorski used perturbation theory to study the asymptotic expansion of exchange energy in H$_2^+$ and discussed a new approach to exchange energy via a variational volume-integral formula.   

The plan of this paper is as follows: in the next section, we derive the arbitrary-spin first-order exchange energy formula for two monomers described by their spin-restricted determinants and show the connection with the Heisenberg model of the scalar interaction of two spins. 
In Section III we give the most important details regarding the implementation of the theory, and in Section IV we report the test calculations for several important open-shell$\cdots$open-shell systems.  In the last section of the paper, we  summarize our results and discuss prospects for the future.

\section{Theory}
\label{theory}
Similar to the existing closed-shell and high-spin open-shell SAPT
approaches, the permutational symmetry adaptation, leading to the
exchange corrections, is performed in the simplest manner, within the
symmetrized Rayleigh-Schr\"{o}dinger (SRS) 
formalism \cite{Jeziorski:78,Jeziorski:94}. Specifically, the
SRS wavefunction corrections are computed from the standard
Rayleigh-Schr\"{o}dinger (RS) perturbation expansion without any symmetry adaptation:
only later, the energy corrections are computed from a formula that
involves a symmetry projector. In the spin-free formalism used, for example,
in Ref.~\cite{Patkowski:01}, this symmetry operator projects onto a 
particular irreducible representation of the permutation group
$S_{N_A+N_B}$ ($N_X$ is the number of electrons of molecule $X$): the choice 
of this representation is determined by the spin multiplicity. In the more
conventional, spin formalism employed here (and used in the high-spin case
so far), this operator has to be replaced by the $(N_A+N_B)$-electron antisymmetrizer 
and an appropriate spin projection.
In the already existing high-spin open-shell
SAPT theories the monomer wavefunctions $\Psi_A$ and $\Psi_B$
as well as the zeroth-order dimer
wavefunction $\Psi_0=\Psi_A\Psi_B$ are pure spin functions
so the spin projection is not needed. This is only the case if all unpaired
electrons on both monomers have the same spin, that is, when 
$M_{S}=\pm (S_A+S_B)$ so that no contamination by lower-spin states is
possible. Indeed, the restricted open-shell Hartree-Fock and Kohn-Sham 
(ROHF/ROKS) SAPT exchange corrections
have been developed \cite{Zuchowski:08} under the assumption that all unpaired
spins point the same way.

For the low-spin case, it is known \cite{Patkowski:02,Jeziorski:10} that
the antisymmetrizer ${\cal A}$ in the SRS energy expression has to be accompanied
by an operator $P_{SM_S}$ that projects a dimer function onto
a subspace corresponding to a particular value of the total spin $S$ (with an
appropriate $M_S$). In this way,
the exchange corrections, different for different dimer spin
states, are obtained by expanding the energy expression
\begin{equation}\label{estart}
E_{\rm int}=\frac{\langle \Psi_A\Psi_B|V{\cal A}P_{SM_S}|\Phi_{AB}\rangle}{%
\langle \Psi_A\Psi_B|{\cal A}P_{SM_S}|\Phi_{AB}\rangle},
\end{equation}
in powers of the intermolecular interaction operator $V$ and separating the
polarization and exchange effects in each order. In Eq.~(\ref{estart}),
${\cal A}$ is the $(N_A+N_B)$-electron antisymmetrizer and 
$\Phi_{AB}$ is the polarization (RS) expansion of the wave function
for the complex \cite{Jeziorski:94}. We will
assume that $\Psi_A$ and $\Psi_B$ are ground-state high-spin open-shell determinants.
The occupied orbitals in $\Psi_A$ are $\chi_{i_1}$, $\chi_{i_2}$, \ldots,
$\chi_{i_{k_A}}$ (inactive, doubly occupied) and $\chi_{a_1}$, $\chi_{a_2}$, 
\ldots, $\chi_{a_{2S_A}}$ (active, occupied by $\alpha$ electrons only).
The occupied orbitals in $\Psi_B$ are $\chi_{j_1}$, $\chi_{j_2}$, \ldots,
$\chi_{j_{k_B}}$ (inactive, doubly occupied) and $\chi_{b_1}$, $\chi_{b_2}$, 
\ldots, $\chi_{b_{2S_B}}$ (active, occupied by $\beta$ electrons only).

The spin projector $P_{SM_S}$ commutes with $V$ and the zeroth-order Hamiltonian
$H_0=H_A+H_B$ as the latter operators are spin-independent. Moreover, the operator
$\hat{S}^2$, and thus also $P_{SM_S}$, is independent of the ordering of electrons and
hence commutes with any electron permutations or their linear combinations
including ${\cal A}$. As a result, we can rewrite Eq.~(\ref{estart}) in an
alternative form where the spin projection acts in the bra instead of the ket:
\begin{equation}\label{estart_alt}
E_{\rm int}=\frac{\langle P_{SM_S}\Psi_A\Psi_B|V{\cal A}|\Phi'_{AB}\rangle}{%
\langle P_{SM_S}\Psi_A\Psi_B|{\cal A}|\Phi'_{AB}\rangle}.
\end{equation}
While Eqs.~(\ref{estart}) and (\ref{estart_alt}) are fully equivalent, they have 
quite different interpretations.
Eq.~(\ref{estart}) corresponds to a standard RS treatment where the wavefunction
corrections in the expansion of $\Phi_{AB}$ are the same for all asymptotically 
degenerate multiplets: the spin
projection enters later, at the calculation of SRS energy corrections. 
Eq.~(\ref{estart_alt}) corresponds to a perturbation expansion initiated from a
spin-adapted zeroth-order function $P_{SM_S}\Psi_A\Psi_B$, with all resulting 
perturbation corrections (in the expansion of $\Phi'_{AB}$) also being of pure spin. 
Importantly, both approaches
allow the use of standard nondegenerate perturbation theory 
(unless degeneracy arises from the orbital part of the monomer wavefunctions,
which is outside the scope of this work), albeit for quite different reasons.
For Eq.~(\ref{estart_alt}), the projector $P_{SM_S}$ reduces the zeroth-order
space of $(2S_A+1)(2S_B+1)$ asymptotically degenerate 
product functions corresponding to a pair of
interacting multiplets to a single spin-adapted combination with the requested
$(S,M_S)$. For Eq.~(\ref{estart}), any of the $(2S_A+1)(2S_B+1)$ product 
functions, including $\Psi_A\Psi_B$, is a valid zeroth-order state for nondegenerate RS perturbation
theory as neither $V$ nor $H_0$ mix states with different $M_{S_A}$ or $M_{S_B}$.
We choose Eq.~(\ref{estart}) as the starting point for the derivation below.


We now introduce the single exchange approximation (often called the $S^2$ approximation
as it neglects terms higher than quadratic in the intermolecular overlap integrals):
\begin{equation}\label{singex}
{\cal A}\sim (1+{\cal P}){\cal A}_A {\cal A}_B,
\end{equation}
where the single-exchange operator ${\cal P}$ is given by
\begin{equation}
{\cal P}=-\sum_{i\in A}\sum_{j\in B} {\cal P}_{ij}
\end{equation}
(for non-approximated alternatives in the closed-shell
case, see Refs.~\cite{Jeziorski:76,Schaffer:12,Schaffer:13}).
We now multiply both sides of Eq. (\ref{estart}) by the denominator,
keeping only the terms of first order in $V$ (thus, replacing
$\Phi_{AB}$ by its zeroth-order term $\Psi_A\Psi_B$), and make
the approximation (\ref{singex}). Introducing the shorthand
notation $\langle X\rangle\equiv\langle \Psi_A\Psi_B|X|\Psi_A\Psi_B\rangle$, we obtain (note that ${\cal A}$, 
${\cal A}_A$, and ${\cal A}_B$ all commute with $P_{SM_S}$)
\begin{equation}\label{est2}
\left(\langle V\rangle+E^{(10)}_{\rm exch}\right) 
\left(\langle P_{SM_S}\rangle
+\langle {\cal P}P_{SM_S}\rangle\right)
=\langle VP_{SM_S}\rangle
+\langle V{\cal P}P_{SM_S}\rangle
\end{equation}
Note that we have separated the electrostatic contribution 
$E^{(10)}_{\rm elst}=\langle V\rangle$ from the exchange one, and
we introduced the zero into the $E^{(10)}$ superscript to remind the
reader that no intramolecular correlation effects are included.

Before we proceed, we need to understand a bit better the action of the 
operator $P_{SM_S}$ on the product $\Psi_A\Psi_B$. The RHF determinants $\Psi_A$
and $\Psi_B$ are pure spin functions corresponding to the quantum numbers
$(S_A,M_{S_A}=S_A)$ and $(S_B,M_{S_B}=-S_B)$, respectively.
In the zeroth-order space spanned by $(2S_A+1)(2S_B+1)$ product functions that correspond 
to total spin $S_A$ for molecule A and total spin $S_B$ for molecule B (and any 
combination of $M_{S_A},M_{S_B}$), there exists exactly one function $\Psi_{S,M_S}$ 
with a total spin $S\in\{|S_A-S_B|,\ldots,S_A+S_B\}$ and its projection 
$M_S\in\{-S,\ldots,S\}$. This function is a linear combination of only those 
product functions that correspond to $M_{S_A}+M_{S_B}=M_S$. Therefore, the 
projection $P_{SM_S}\Psi_A\Psi_B$ picks, up to normalization, this very 
function $\Psi_{S,M_S}$ out of the zeroth-order space. 
In other words, $P_{SM_S}\Psi_A\Psi_B$
produces a linear combination of products of functions of A and B with the
same values $S_A,S_B$ but different $M_{S_A},M_{S_B}$ (however, the latter two 
numbers add up to the same total $M_S$). The coefficients in this linear
combination are the Clebsch-Gordan coefficients 
$\langle SM_S | S_A M_{S_A} S_B M_{S_B}\rangle$ --- we will denote the
coefficients
$\langle S\: (S_A-S_B) | S_A\: S_A\: S_B\: -S_B\rangle$,
$\langle S\: (S_A-S_B) | S_A\: (S_A-1)\: S_B\: (-S_B+1)\rangle$,
and $\langle S\: (S_A-S_B) | S_A\: (S_A-2)\: S_B\: (-S_B+2)\rangle$ by
$c_0$, $c_1$, and $c_2$, respectively.
Specifically, the spin
projection can be expressed as follows:
\begin{equation}\label{pspsi0}
P_{SM_S}\Psi_A\Psi_B=c_0 \Psi_A\Psi_B
+c_1 
\Psi_A^{\downarrow}\Psi_B^{\uparrow}
+c_2
\Psi_A^{\downarrow\downarrow}\Psi_B^{\uparrow\uparrow}+\ldots
\end{equation}
where, for example, $\Psi_A^{\downarrow}$ is a normalized wavefunction
(linear combination of determinants) corresponding to the spin quantum
numbers $(S_A,S_A-1)$. Up to a constant, this function can be obtained from
$\Psi_A$ by the action of the spin-lowering operator $\hat{S}_{-}$.

Eq. (\ref{pspsi0}) implies that
\begin{equation}
\langle P_{SM_S}\rangle=c_0
\end{equation}
\begin{equation}
\langle VP_{SM_S}\rangle=c_0 \langle V\rangle
\end{equation}
Therefore, Eq. (\ref{est2}) becomes
\begin{equation}\label{est3}
\langle V\rangle\langle {\cal P}P_{SM_S}\rangle
+c_0 E^{(10)}_{\rm exch}
+E^{(10)}_{\rm exch} 
\langle {\cal P}P_{SM_S}\rangle
=
\langle V{\cal P}P_{SM_S}\rangle
\end{equation}
Now, the last term on the l.h.s. is neglected as it requires at least two
electron exchanges (one in $E^{(10)}_{\rm exch}$ and one in $\langle {\cal P}P_{SM_S}\rangle$), 
and we arrive at the final formula for the SRS-like
first-order exchange energy:
\begin{widetext}
\begin{equation}\label{e1exch}
E^{(10)}_{\rm exch} 
=c_0^{-1}\left(\langle V{\cal P}P_{SM_S}\rangle
-\langle V\rangle
\langle {\cal P}P_{SM_S}\rangle\right)
\end{equation}
\end{widetext}
Let us now go back to Eq. (\ref{pspsi0}) and examine the spin-flipped
monomer wavefunctions such as $\Psi_A^{\downarrow}$. Up to normalization, this
function is equal to $\hat{S}_{-}\Psi_A$, where $\hat{S}_{-}=\sum_{k=1}^{N_A} \hat{S}_{-}(k)$
applies the conventional spin-lowering operator 
$\hat{S}_{-}(k)=\hat{S}_{x}(k)-i\hat{S}_{y}(k)$ to
each of the $N_A$ spins. It can be shown that when $\Psi_A$
is a restricted high-spin determinant like in this work, the lowering
operator acts on it as follows,
\begin{widetext}
\begin{equation}\label{sminuspsi}
\hat{S}_{-}\Psi_A=\sum_{n=1}^{2S_A} \left|\chi_{i_1}\alpha \chi_{i_1}\beta \ldots
\chi_{i_{k_A}}\alpha \chi_{i_{k_A}}\beta \chi_{a_1}\alpha \ldots 
\chi_{a_{n-1}}\alpha
\chi_{a_n}\beta \chi_{a_{n+1}}\alpha \ldots \chi_{a_{2S_A}}\alpha \right|
\end{equation}
\end{widetext}
where the spin orbitals present in each (normalized) determinant are
explicitly listed. Note that only the active spin orbitals end up being
spin-flipped. This observation is true for the ROHF determinants only:
lowering of the spin for an inactive spin orbital results in a duplicate spin
orbital and the determinant vanishes. If UHF determinants were considered,
this simplification would not take place. Analogously, the repeated application
of the spin-lowering operator, $\hat{S}_{-}\hat{S}_{-}\Psi_A$, gives a linear combination
of all possible determinants where two active spinorbitals have been flipped
from $\alpha$ to $\beta$. The function $\hat{S}_{-}\Psi_A$, Eq. (\ref{sminuspsi}), is
not normalized, but all $2S_A$ determinants entering the linear combination are
clearly orthonormal. Therefore, $\Psi_A^{\downarrow}=(1/\sqrt{2S_A}) \hat{S}_{-}\Psi_A$
is normalized.

Inserting Eq. (\ref{pspsi0}) into Eq. (\ref{e1exch}), we obtain
\begin{widetext}
\begin{eqnarray}\label{e1excha}
E^{(10)}_{\rm exch} 
&=&
\langle V{\cal P}\rangle
+c_1/c_0\langle \Psi_A\Psi_B|V{\cal P}|\Psi_A^{\downarrow}\Psi_B^{\uparrow}\rangle
\nonumber \\ &&
+c_2/c_0\langle \Psi_A\Psi_B|V{\cal P}|
\Psi_A^{\downarrow\downarrow}\Psi_B^{\uparrow\uparrow}\rangle
+\ldots+
\nonumber \\ &&
-\langle V\rangle \left(
\langle {\cal P}\rangle
+c_1/c_0\langle \Psi_A\Psi_B|{\cal P}|\Psi_A^{\downarrow}\Psi_B^{\uparrow}\rangle
\right. \nonumber \\ && \left.
+c_2/c_0\langle \Psi_A\Psi_B|{\cal P}|
\Psi_A^{\downarrow\downarrow}\Psi_B^{\uparrow\uparrow}\rangle
+\ldots\right)
\end{eqnarray}
\end{widetext}
The first interesting observation from Eq. (\ref{e1excha}) is that the
standard, closed-shell like contribution (the ``spin-diagonal'' term)
$\langle V{\cal P}\rangle-\langle V\rangle\langle {\cal P}\rangle$
is present in the first-order exchange energy of any dimer spin state
regardless of the values of the Clebsch-Gordan coefficients. It is the
off-diagonal ``spin-flip'' terms that are responsible for the splittings (and, as we will
see below, only the single spin-flip term survives). Starting with the
diagonal term and using the density-matrix approach to exchange energy that
is valid in monomer- as well as dimer-centered basis sets, we can use 
the general spinorbital
expression given by Eq. (39) of Moszy\'nski et al. \cite{Moszynski:94a}:
\begin{widetext}
\begin{eqnarray}\label{e1exmosz}
\langle V{\cal P}\rangle -\langle V\rangle\langle {\cal P}\rangle &=&
-\left[ \tilde{v}^{\beta\alpha}_{\alpha\beta}
+S^{\beta}_{\alpha'}(\tilde{v}^{\alpha\alpha'}_{\alpha\beta}
-\tilde{v}^{\alpha'\alpha}_{\alpha\beta})
+S^{\alpha}_{\beta'}(\tilde{v}^{\beta'\beta}_{\alpha\beta}
-\tilde{v}^{\beta\beta'}_{\alpha\beta})
\right. \nonumber \\ && \left.
-S^{\beta}_{\alpha'}S^{\alpha'}_{\beta'}\tilde{v}^{\alpha\beta'}_{\alpha\beta}
-S^{\beta'}_{\alpha'}S^{\alpha}_{\beta'}\tilde{v}^{\alpha'\beta}_{\alpha\beta}
+S^{\beta}_{\alpha'}S^{\alpha}_{\beta'}\tilde{v}^{\alpha'\beta'}_{\alpha\beta}
\right]
\end{eqnarray}
\end{widetext}
In that reference, $\alpha,\alpha'$ ($\beta,\beta'$) run over all occupied
spin orbitals of A (B). Equation (\ref{e1exmosz}) applies in the open-shell
case as well (for the diagonal term) but we have to break up the summation
over e.g. the occupied spin orbitals of A into the inactive orbitals
with spins $\alpha$ and $\beta$ and active orbitals with spin $\alpha$.
When we do that and perform all spin integrations, the resulting expression
for the diagonal term is
\begin{widetext}
\begin{eqnarray}\label{e1exdiag}
E^{(10)}_{\rm exch,diag}=\langle V{\cal P}\rangle -\langle V\rangle\langle {\cal P}\rangle &=&
- \tilde{v}^{ja}_{aj}
- \tilde{v}^{bi}_{ib}
-2 \tilde{v}^{ji}_{ij}
\nonumber \\ &&
+ \tilde{v}^{a'a}_{aj} S^{j}_{a'}
- \tilde{v}^{aa'}_{aj} S^{j}_{a'}
+ \tilde{v}^{jj'}_{aj} S^{a}_{j'}
-2 \tilde{v}^{j'j}_{aj} S^{a}_{j'}
+ \tilde{v}^{ai}_{ij} S^{j}_{a}
-2 \tilde{v}^{ia}_{ij} S^{j}_{a}
- \tilde{v}^{jb}_{ab} S^{a}_{j}
\nonumber \\ &&
+ \tilde{v}^{bb'}_{ib} S^{i}_{b'}
- \tilde{v}^{b'b}_{ib} S^{i}_{b'}
+ \tilde{v}^{i'i}_{ib} S^{b}_{i'}
-2 \tilde{v}^{ii'}_{ib} S^{b}_{i'}
+ \tilde{v}^{jb}_{ij} S^{i}_{b}
-2 \tilde{v}^{bj}_{ij} S^{i}_{b}
- \tilde{v}^{ai}_{ab} S^{b}_{i}
\nonumber \\ &&
+ \tilde{v}^{ia}_{aj} S^{j}_{i}
+ \tilde{v}^{bj}_{ib} S^{i}_{j}
-2 \tilde{v}^{ai}_{aj} S^{j}_{i}
-2 \tilde{v}^{jb}_{ib} S^{i}_{j}
\nonumber \\ &&
+2 \tilde{v}^{i'i}_{ij} S^{j}_{i'}
+2 \tilde{v}^{jj'}_{ij} S^{i}_{j'}
-4 \tilde{v}^{ii'}_{ij} S^{j}_{i'}
-4 \tilde{v}^{j'j}_{ij} S^{i}_{j'}
\nonumber \\ &&
+ \tilde{v}^{a'b}_{ab} S^{j}_{a'} S^{a}_{j}
+ \tilde{v}^{ab'}_{ab} S^{b}_{i} S^{i}_{b'}
+2 \tilde{v}^{ab}_{ib} S^{j}_{a} S^{i}_{j}
+2 \tilde{v}^{ab}_{aj} S^{j}_{i} S^{i}_{b}
\nonumber \\ &&
+2 \tilde{v}^{a'j}_{aj} S^{j'}_{a'} S^{a}_{j'}
+2 \tilde{v}^{aj'}_{aj} S^{j}_{i} S^{i}_{j'}
+ \tilde{v}^{aj'}_{aj} S^{j}_{a'} S^{a'}_{j'}
- \tilde{v}^{a'j'}_{aj} S^{j}_{a'} S^{a}_{j'}
\nonumber \\ &&
+2 \tilde{v}^{ib'}_{ib} S^{b}_{i'} S^{i'}_{b'}
+2 \tilde{v}^{i'b}_{ib} S^{j}_{i'} S^{i}_{j}
+ \tilde{v}^{i'b}_{ib} S^{b'}_{i'} S^{i}_{b'}
- \tilde{v}^{i'b'}_{ib} S^{b}_{i'} S^{i}_{b'}
\nonumber \\ &&
+4 \tilde{v}^{aj}_{ij} S^{j'}_{a} S^{i}_{j'}
-2 \tilde{v}^{aj'}_{ij} S^{j}_{a} S^{i}_{j'}
+4 \tilde{v}^{ib}_{ij} S^{j}_{i'} S^{i'}_{b}
-2 \tilde{v}^{i'b}_{ij} S^{j}_{i'} S^{i}_{b}
\nonumber \\ &&
+2 \tilde{v}^{i'j}_{ij} S^{b}_{i'} S^{i}_{b}
+4 \tilde{v}^{i'j}_{ij} S^{j'}_{i'} S^{i}_{j'}
+2 \tilde{v}^{ij'}_{ij} S^{j}_{a} S^{a}_{j'}
+4 \tilde{v}^{ij'}_{ij} S^{j}_{i'} S^{i'}_{j'}
\nonumber \\ &&
-2 \tilde{v}^{i'j'}_{ij} S^{j}_{i'} S^{i}_{j'}
\end{eqnarray}
\end{widetext}
where the implicit summations now run over orbitals: $i$--(inactive A),
$j$--(inactive B), $a$--(active A), $b$--(active B). It is worth noting
that an active-active exchange term $\tilde{v}^{ba}_{ab}$ is absent from
Eq. (\ref{e1exdiag}) due to the fact that all active orbitals on A are
paired with $\alpha$ spins and all active orbitals on B are paired with
$\beta$ spins. 

For the first off-diagonal term, corresponding to a single spin flip between
the interacting monomers, Eq. (\ref{e1excha}) involves two matrix
elements, $\langle \Psi_A\Psi_B|V{\cal P}|\Psi_A^{\downarrow}\Psi_B^{\uparrow}\rangle$
and $\langle \Psi_A\Psi_B|{\cal P}|\Psi_A^{\downarrow}\Psi_B^{\uparrow}\rangle$.
Analogously to the conventional closed-shell formalism, these elements
can be expressed through the ``spin-flip interaction density matrix'':
\begin{equation}
\langle \Psi_A\Psi_B|V{\cal P}|\Psi_A^{\downarrow}\Psi_B^{\uparrow}\rangle=
\int \tilde{v}(12) \rho_{int}^{\downarrow\uparrow}(12)\,{\rm d}\tau_{12}
\end{equation}
\begin{equation}\label{ren1sf}
\langle \Psi_A\Psi_B|{\cal P}|\Psi_A^{\downarrow}\Psi_B^{\uparrow}\rangle=
\frac{1}{N_A N_B} \int \rho_{int}^{\downarrow\uparrow}(12)\,{\rm d}\tau_{12}
\end{equation}
where
\begin{equation}
\rho_{int}^{\downarrow\uparrow}(12)=N_A N_B \int \Psi_A\Psi_B {\cal P}
\Psi_A^{\downarrow}\Psi_B^{\uparrow}\,{\rm d}\tau'_{12}
\end{equation}
and, as always, ${\rm d}\tau'_{12}$ means the integration over coordinates of
all electrons except 1 and 2. The spin-flip interaction density matrix can
be expressed by the (also spin-flip) one- and two-electron reduced density
matrices in full analogy to Eq. (99) of Ref.~\cite{Patkowski:06}:
\begin{widetext}
\begin{eqnarray}\label{rhointsf}
\rho_{int}^{\downarrow\uparrow}(12)&=&
-\rho_A^{\downarrow}(1|2)\rho_B^{\uparrow}(2|1)
\nonumber \\ &&
-\int \rho_A^{\downarrow}(1|4)\Gamma_B^{\uparrow}(24|21)\,{\rm d}\tau_4
\nonumber \\ &&
-\int \Gamma_A^{\downarrow}(13|12)\rho_B^{\uparrow}(2|3)\,{\rm d}\tau_3
\nonumber \\ &&
-\int \Gamma_A^{\downarrow}(13|14)\Gamma_B^{\uparrow}(24|23)\,{\rm d}\tau_3
{\rm d}\tau_4
\end{eqnarray}
\end{widetext}
The spin-flip reduced density matrices on the r.h.s. of Eq.~(\ref{rhointsf}) are 
in fact no different from ordinary transition density matrices: for example, for
monomer A
\begin{equation}\label{sf1rdm}
\rho_A^{\downarrow}(1|1')=N_A\int \Psi_A^{\ast}(1,2,\ldots,N_A)\Psi_A^{\downarrow}(1',2,\ldots,N_A)
{\rm d}\tau'_1
\end{equation}
\begin{equation}\label{sf2rdm}
\Gamma_A^{\downarrow}(12|1'2')=N_A(N_A-1)
\int \Psi_A^{\ast}(1,2,3,\ldots,N_A)\Psi_A^{\downarrow}(1',2',3,\ldots,N_A) {\rm d}\tau'_{12}
\end{equation}
Therefore, in order to obtain spinorbital expressions for these matrices, we
can use Eqs. (92) and (94) of Ref.~\cite{Patkowski:06},
except that a general $\alpha\to\rho$ excitation is replaced by a linear
combination of spin-flip
excitations. Therefore, we can straightforwardly use Eq. (92) 
of Ref.~\cite{Patkowski:06} to write
\begin{equation}
\rho_A^{\downarrow}(1|1')= \frac{1}{\sqrt{2S_A}} \sum_{n=1}^{2S_A}
\chi_{a_n}\alpha(1)\chi_{a_n}\beta(1')
\end{equation}
The application of Eq. (94) from Ref.~\cite{Patkowski:06} 
to obtain a formula for $\Gamma_A^{\downarrow}$
is a little more complicated because there is an additional summation over
an occupied orbital in this equation. This summation needs to be split
into three: over inactive orbitals with spin $\alpha$, over inactive orbitals
with spin $\beta$, and over active orbitals (which have spin $\alpha$). The
resulting expression is
\begin{widetext}
\begin{eqnarray}
\Gamma_A^{\downarrow}(12|1'2')&=&\frac{1}{\sqrt{2S_A}}\sum_{n=1}^{2S_A}\left[
\sum_{m=1}^{k_A} \left( 
\chi_{a_n}\alpha(1)\chi_{i_m}\alpha(2)\chi_{a_n}\beta(1')\chi_{i_m}\alpha(2')
\right. \right. \nonumber\\ && \left.\left.
-\chi_{i_m}\alpha(1)\chi_{a_n}\alpha(2)\chi_{a_n}\beta(1')\chi_{i_m}\alpha(2')
\right. \right. \nonumber\\ && \left.\left.
-\chi_{a_n}\alpha(1)\chi_{i_m}\alpha(2)\chi_{i_m}\alpha(1')\chi_{a_n}\beta(2')
\right. \right. \nonumber\\ && \left.\left.
+\chi_{i_m}\alpha(1)\chi_{a_n}\alpha(2)\chi_{i_m}\alpha(1')\chi_{a_n}\beta(2')
\right)
\right. \nonumber\\ && \left.
+\sum_{m=1}^{k_A} \left(
\chi_{a_n}\alpha(1)\chi_{i_m}\beta(2)\chi_{a_n}\beta(1')\chi_{i_m}\beta(2')
\right. \right. \nonumber\\ && \left.\left.
-\chi_{i_m}\beta(1)\chi_{a_n}\alpha(2)\chi_{a_n}\beta(1')\chi_{i_m}\beta(2')
\right. \right. \nonumber\\ && \left.\left.
-\chi_{a_n}\alpha(1)\chi_{i_m}\beta(2)\chi_{i_m}\beta(1')\chi_{a_n}\beta(2')
\right. \right. \nonumber\\ && \left.\left.
+\chi_{i_m}\beta(1)\chi_{a_n}\alpha(2)\chi_{i_m}\beta(1')\chi_{a_n}\beta(2')
\right)
\right. \nonumber\\ && \left.
+\sum_{m=1}^{2S_A} \left(
\chi_{a_n}\alpha(1)\chi_{a_m}\alpha(2)\chi_{a_n}\beta(1')\chi_{a_m}\alpha(2')
\right. \right. \nonumber\\ && \left.\left.
-\chi_{a_m}\alpha(1)\chi_{a_n}\alpha(2)\chi_{a_n}\beta(1')\chi_{a_m}\alpha(2')
\right. \right. \nonumber\\ && \left.\left.
-\chi_{a_n}\alpha(1)\chi_{a_m}\alpha(2)\chi_{a_m}\alpha(1')\chi_{a_n}\beta(2')
\right. \right. \nonumber\\ && \left.\left.
+\chi_{a_m}\alpha(1)\chi_{a_n}\alpha(2)\chi_{a_m}\alpha(1')\chi_{a_n}\beta(2')
\right)\raisebox{0pt}[10pt][10pt]{}
\right]
\end{eqnarray}
\end{widetext}
Developing analogous formulas for the spin-flip 
reduced density matrices of monomer
B, employing Eq.~(\ref{rhointsf}), and performing spin integration, one
arrives at the following formula:
\begin{widetext}
\begin{eqnarray}\label{offdiag1}
\langle \Psi_A\Psi_B|V{\cal P}|\Psi_A^{\downarrow}\Psi_B^{\uparrow}\rangle&=&
\frac{1}{2\sqrt{S_A S_B}}\left[
- \tilde{v}_{ab}^{ba} \right.
\nonumber \\ &&
+ \tilde{v}_{ab}^{bj} S_{j}^{a}
+ \tilde{v}_{aj}^{jb} S_{b}^{a}
-2 \tilde{v}_{aj}^{bj} S_{b}^{a}
+ \tilde{v}_{ab}^{bb'} S_{b'}^{a}
- \tilde{v}_{ab'}^{bb'} S_{b}^{a}
\nonumber \\ &&
+ \tilde{v}_{ab}^{ia} S_{i}^{b}
+ \tilde{v}_{ib}^{ai} S_{a}^{b}
-2 \tilde{v}_{ia}^{ib} S_{b}^{a}
+ \tilde{v}_{ab}^{a'a} S_{a'}^{b}
- \tilde{v}_{a'a}^{a'b} S_{b}^{a}
\nonumber \\ &&
-2 \tilde{v}_{ab}^{ij} S_{i}^{b} S_{j}^{a}
-2 \tilde{v}_{aj}^{ib} S_{i}^{j} S_{b}^{a}
+4 \tilde{v}_{aj}^{ij} S_{i}^{b} S_{b}^{a}
+4 \tilde{v}_{ib}^{ij} S_{a}^{b} S_{j}^{a}
-4 \tilde{v}_{ij}^{ij} S_{a}^{b} S_{b}^{a}
\nonumber \\ &&
+2 \tilde{v}_{ib}^{ib'} S_{a}^{b} S_{b'}^{a}
-2 \tilde{v}_{ab}^{ib'} S_{i}^{b} S_{b'}^{a}
-2 \tilde{v}_{ib'}^{ib'} S_{a}^{b} S_{b}^{a}
+2 \tilde{v}_{ib'}^{ab'} S_{a}^{b} S_{b}^{i}
\nonumber \\ &&
+2 \tilde{v}_{aj}^{a'j} S_{a'}^{b} S_{b}^{a}
-2 \tilde{v}_{ab}^{a'j} S_{a'}^{b} S_{j}^{a}
-2 \tilde{v}_{a'j}^{a'j} S_{a}^{b} S_{b}^{a}
+2 \tilde{v}_{a'b}^{a'j} S_{a}^{b} S_{j}^{a}
\nonumber \\ && \left.
+ \tilde{v}_{a'b}^{a'b'} S_{a}^{b} S_{b'}^{a}
- \tilde{v}_{a'b'}^{a'b'} S_{a}^{b} S_{b}^{a}
- \tilde{v}_{ab}^{a'b'} S_{a'}^{b} S_{b'}^{a}
+ \tilde{v}_{ab'}^{a'b'} S_{a'}^{b} S_{b}^{a}  \right]
\end{eqnarray}
\end{widetext}
For the renormalization term, Eq.~(\ref{ren1sf}), we find (for example,
by reanalyzing Eq.~(\ref{offdiag1}), replacing each $\tilde{v}$ by
$\frac{1}{N_A N_B}$ times the appropriate product of overlap integrals) that
\begin{equation}\label{ren1a}
\langle \Psi_A\Psi_B|{\cal P}|\Psi_A^{\downarrow}\Psi_B^{\uparrow}\rangle=
-\frac{1}{2\sqrt{S_A S_B}}S_{a}^{b} S_{b}^{a}
\end{equation}
The ROHF electrostatic energy is equal to
\begin{equation}\label{ren1b}
E^{(10)}_{\rm elst}=\langle V\rangle=
4\tilde{v}^{ij}_{ij}+2\tilde{v}^{aj}_{aj}+2\tilde{v}^{ib}_{ib}
+\tilde{v}^{ab}_{ab}
\end{equation}
Combining Eqs.~(\ref{offdiag1})--(\ref{ren1b}), we obtain the final formula
for the single spin-flip contribution to Eq.~(\ref{e1excha}):
\begin{widetext}
\begin{eqnarray}\label{offdiag1final}
\lefteqn{\frac{1}{2\sqrt{S_A S_B}} E^{(10)}_{\rm exch,1flip}=
\langle \Psi_A\Psi_B|V{\cal P}|\Psi_A^{\downarrow}\Psi_B^{\uparrow}\rangle
-\langle V\rangle
\langle \Psi_A\Psi_B|{\cal P}|\Psi_A^{\downarrow}\Psi_B^{\uparrow}\rangle=}
  \\ &&
\frac{1}{2\sqrt{S_A S_B}}\left[ - \tilde{v}_{ab}^{ba} \right.  
+ \tilde{v}_{ab}^{bj} S_{j}^{a}
+ \tilde{v}_{aj}^{jb} S_{b}^{a}
-2 \tilde{v}_{aj}^{bj} S_{b}^{a}
+ \tilde{v}_{ab}^{bb'} S_{b'}^{a}
- \tilde{v}_{ab'}^{bb'} S_{b}^{a}
\nonumber \\ && 
+ \tilde{v}_{ab}^{ia} S_{i}^{b}
+ \tilde{v}_{ib}^{ai} S_{a}^{b}
-2 \tilde{v}_{ia}^{ib} S_{b}^{a}
+ \tilde{v}_{ab}^{a'a} S_{a'}^{b}
- \tilde{v}_{a'a}^{a'b} S_{b}^{a}
\nonumber \\ &&
-2 \tilde{v}_{ab}^{ij} S_{i}^{b} S_{j}^{a}
-2 \tilde{v}_{aj}^{ib} S_{i}^{j} S_{b}^{a}
+4 \tilde{v}_{aj}^{ij} S_{i}^{b} S_{b}^{a}
+4 \tilde{v}_{ib}^{ij} S_{a}^{b} S_{j}^{a}
\nonumber  
+2 \tilde{v}_{ib}^{ib'} S_{a}^{b} S_{b'}^{a}
-2 \tilde{v}_{ab}^{ib'} S_{i}^{b} S_{b'}^{a}
+2 \tilde{v}_{ib'}^{ab'} S_{a}^{b} S_{b}^{i}
\nonumber \\ && 
+2 \tilde{v}_{aj}^{a'j} S_{a'}^{b} S_{b}^{a}
-2 \tilde{v}_{ab}^{a'j} S_{a'}^{b} S_{j}^{a}
+2 \tilde{v}_{a'b}^{a'j} S_{a}^{b} S_{j}^{a}
\nonumber  \left.
+ \tilde{v}_{a'b}^{a'b'} S_{a}^{b} S_{b'}^{a}
- \tilde{v}_{ab}^{a'b'} S_{a'}^{b} S_{b'}^{a}
+ \tilde{v}_{ab'}^{a'b'} S_{a'}^{b} S_{b}^{a} \right] 
\end{eqnarray}
\end{widetext}

Finally, we will show that the double spin-flip term in Eq. (\ref{e1excha}),
and all subsequent terms, vanish identically. Specifically, we will
prove that
$\langle \Psi_A\Psi_B|V{\cal P}|
\Psi_A^{\downarrow\downarrow}\Psi_B^{\uparrow\uparrow}\rangle=0$, where
$\Psi_A^{\downarrow\downarrow}$ and $\Psi_B^{\uparrow\uparrow}$ are 
proportional to $\hat{S}_{-}\hat{S}_{-}\Psi_A$ and $\hat{S}_{+}\hat{S}_{+}\Psi_B$, respectively. 
Similar to the previous term,
\begin{equation}\label{offdiag2sf}
\langle \Psi_A\Psi_B|V{\cal P}|\Psi_A^{\downarrow\downarrow}
\Psi_B^{\uparrow\uparrow}\rangle=
\int \tilde{v}(12) \rho_{int}^{\downarrow\downarrow\uparrow\uparrow}(12)
\,{\rm d}\tau_{12}
\end{equation}
where the double spin-flip interaction density matrix 
$\rho_{int}^{\downarrow\downarrow\uparrow\uparrow}(12)$ is given by a formula
analogous to Eq.~(\ref{rhointsf}). However, as a double spin flip can be 
treated as a special case of a double excitation, the one- and two-electron
reduced spin-flip density matrices in the modified Eq.~(\ref{rhointsf}) are
now given by Eqs. (93) and (95), respectively, of Ref.~\cite{Patkowski:06}.
Consequently,
\begin{equation}\label{rho12sf}
\rho_A^{\downarrow\downarrow}(1|1')=0
\end{equation}
and the two-electron density matrix is proportional to
\begin{eqnarray}\label{gamma12sf}
\Gamma_A^{\downarrow\downarrow}(12|1'2')&\sim&\sum_{m,n=1}^{2S_A} 
\left[
\chi_{a_n}\alpha(1)\chi_{a_m}\alpha(2)\chi_{a_n}\beta(1')\chi_{a_m}\beta(2')
\right.  \nonumber\\ && \left.
-\chi_{a_m}\alpha(1)\chi_{a_n}\alpha(2)\chi_{a_n}\beta(1')\chi_{a_m}\beta(2')
\right.  \nonumber\\ && \left.
-\chi_{a_n}\alpha(1)\chi_{a_m}\alpha(2)\chi_{a_m}\beta(1')\chi_{a_n}\beta(2')
\right.  \nonumber\\ && \left.
+\chi_{a_m}\alpha(1)\chi_{a_n}\alpha(2)\chi_{a_m}\beta(1')\chi_{a_n}\beta(2')
\right]
\end{eqnarray}
According to Eq.~(\ref{rho12sf}), only the last term in the expression for
$\rho_{int}^{\downarrow\downarrow\uparrow\uparrow}$ survives:
\begin{equation}
\rho_{int}^{\downarrow\downarrow\uparrow\uparrow}(12)=
-\int \Gamma_A^{\downarrow\downarrow}(13|14)
\Gamma_B^{\uparrow\uparrow}(24|23)\,{\rm d}\tau_3
{\rm d}\tau_4
\end{equation}
Such an expression vanishes upon the integration over ${\rm d}\tau_1$
and ${\rm d}\tau_2$, either alone or with a spin-independent operator such
as $\tilde{v}(12)$ (like in Eq. (\ref{offdiag2sf})). In particular, in the
integration over ${\rm d}\tau_1$, the two spinorbitals in 
$\Gamma_A^{\downarrow\downarrow}(13|14)$ that depend on the coordinates of
electron 1 always occur with opposite spins (cf. Eq. (\ref{gamma12sf})), 
leading to a zero spin integral.
This concludes the proof that the double spin-flip contribution to
Eq. (\ref{e1excha}) vanishes.

To conclude, the first-order SAPT exchange correction for an interaction
of two high-spin open-shell molecules with spins $S_A$ and $S_B$, in an  
arbitrary dimer spin state $|S_A-S_B|\le S\le S_A+S_B$, is given by the formula 
which combines the orbital expressions for $E^{(10)}_{\rm exch,diag}$ and 
$E^{(10)}_{\rm exch,1flip}$ defined in Eqs. (\ref{e1exdiag}) and
(\ref{offdiag1final}), respectively. By simplifying the $c_1/\left(2c_0\sqrt{S_A S_B}\right)$ coefficient,
as outlined in the Appendix, 
we obtain 
\begin{eqnarray}\label{e1exchf}
E^{(10)}_{\rm exch} 
&=&
\langle V{\cal P}\rangle
-\langle V\rangle 
\langle {\cal P}\rangle
\nonumber \\ &&
+c_1/c_0\left[
\langle \Psi_A\Psi_B|V{\cal P}|\Psi_A^{\downarrow}\Psi_B^{\uparrow}\rangle
-\langle V\rangle 
\langle \Psi_A\Psi_B|{\cal P}|\Psi_A^{\downarrow}\Psi_B^{\uparrow}\rangle\right]
\nonumber \\ & =& E^{(10)}_{\rm exch,diag} + Z(S_A,S_B,S)  E^{(10)}_{\rm exch,1flip} 
\end{eqnarray}
where 
\begin{equation}\label{clebsch}
Z(S_A,S_B,S) = \frac{S(S+1)+2S_A S_B-S_A(S_A+1) -S_B(S_B+1)}{ 4 S_A S_B}.
\end{equation}
Note that $Z(S_A,S_B,S_A+S_B)=1$, while $Z(S_A,S_B,|S_A-S_B|)=-\frac{1}{2\max{(S_A,S_B)}}$.
 All of the resulting matrix elements are the same
for the full set of asymptotically degenerate multiplets of the dimer: the
only factors in Eq.~(\ref{e1exchf}) that are different for different spin
states are $Z(S_A,S_B,S)$. Thus, as expected from
the Heisenberg Hamiltonian model, the ratios of splittings between different
multiplets are simple, system-independent numbers  (expressible through
the Clebsch-Gordan coefficients) as long as the single exchange approximation
is applied. The observation that the single exchange approximation implies the
Heisenberg picture is not new: in fact, it dates back to the work of
Matsen {\em et al.} \cite{Matsen:71} on spin-free quantum chemistry.

For practical applications to large systems, it is advantageous to recast
the newly derived $E^{(10)}_{\rm exch}$ expressions from the molecular-orbital
(MO) basis into the atomic orbital (AO) one, similar to the AO first-order
exchange expressions without the single-exchange approximation for
closed-shell SAPT \cite{Hesselmann:05} and UHF-based high-spin open-shell
SAPT \cite{Hapka:12}. The AO approach is advantageous as it avoids the 
AO-MO transformation of many different types of two-electron integrals
present in Eqs. (\ref{e1exdiag}) and (\ref{offdiag1final}). Moreover, as
we will see below, the AO expressions make heavy use of the generalized
Coulomb and exchange operators, whose evaluation in the {\sc psi4} 
code \cite{Parrish:17} is highly optimized both with and without
density fitting (DF).

In the following, we will denote AO indices by capital letters assuming, for
simplicity, that the same AO basis set has been used to expand molecular
orbitals of A and B (according to the so-called dimer-centered basis set
formalism \cite{Williams:95}). The SCF coefficients of the molecular
spinorbital $\lambda$ will be denoted by $C^K_{\lambda}$ --- obviously,
in the ROHF formalism, the SCF 
coefficients are the same for spin $\alpha$ and spin $\beta$, for example, $C^K_{i\alpha}=C^K_{i\beta}$.
We will use boldface letters for matrices, and denote by 
${\mathbf A}\cdot {\mathbf B}=\sum_{KL} {\mathbf A}_{KL}{\mathbf B}_{KL}$ the scalar product of two matrices. 
The inactive and active parts of density matrices for monomer A are ${\mathbf P}^{iA}_{KL}=C^K_{i\alpha}C^L_{i\alpha}=
C^K_{i\beta}C^L_{i\beta}$, ${\mathbf P}^{aA}_{KL}=C^K_{a\alpha}C^L_{a\alpha}$, and similarly for monomer B (note that the active electrons on A all have spin $\alpha$, the active electrons on B all have spin $\beta$). 
Therefore, the
total density matrix is  ${\mathbf P}^{A}=2{\mathbf P}^{iA}+{\mathbf P}^{aA}$.
Now, the 
generalized Coulomb and exchange matrices are defined as
\begin{equation}
{\mathbf J}[{\mathbf X}]_{KL} = (KL|MN) {\mathbf X}_{MN}
\end{equation}
\begin{equation}
{\mathbf K}[{\mathbf X}]_{KL} = (KM|NL) {\mathbf X}_{MN}
\end{equation}
and reduce to standard Coulomb and exchange matrices, or their inactive/active contributions, in special cases, for example,
${\mathbf J}^{iA}={\mathbf J}[{\mathbf P}^{iA}]$, ${\mathbf K}^{aA}={\mathbf K}[{\mathbf P}^{aA}]$, \ldots The electrostatic
potential matrix for monomer A is
\begin{equation}
\bm{\omega}_A={\mathbf v}_A+2{\mathbf J}^{iA}+{\mathbf J}^{aA}
\end{equation}
where ${\mathbf v}_A$ is the matrix of the nuclear attraction operator.
By adding the exchange matrices, we obtain the following spin-dependent Fock matrices ${\mathbf h}$:
\begin{equation}
{\mathbf h}_A^{\alpha}={\mathbf v}_A+2{\mathbf J}^{iA}+{\mathbf J}^{aA}-{\mathbf K}^{iA}-{\mathbf K}^{aA}
\end{equation}
\begin{equation}
{\mathbf h}_A^{\beta}={\mathbf v}_A+2{\mathbf J}^{iA}+{\mathbf J}^{aA}-{\mathbf K}^{iA}
\end{equation}
\begin{equation}
{\mathbf h}_B^{\alpha}={\mathbf v}_B+2{\mathbf J}^{iB}+{\mathbf J}^{aB}-{\mathbf K}^{iB}
\end{equation}
\begin{equation}
{\mathbf h}_B^{\beta}={\mathbf v}_B+2{\mathbf J}^{iB}+{\mathbf J}^{aB}-{\mathbf K}^{iB}-{\mathbf K}^{aB}
\end{equation}
With this notation, the diagonal and single spin-flip terms of the ROHF-based $E^{(10)}_{\rm exch}$ formula,
Eqs.~(\ref{e1exdiag}) and (\ref{offdiag1final}), respectively, can be rewritten as follows:
\begin{align}
E^{(10)}_{\rm exch,diag}=&
- {\mathbf P}^{iB} \cdot (2{\mathbf K}^{iA}+{\mathbf K}^{aA})
- {\mathbf P}^{aB} \cdot {\mathbf K}^{iA}
\nonumber \\ &
- ({\mathbf P}^{iA} {\mathbf S}^{AO} {\mathbf P}^{iB}) \cdot ({\mathbf h}_A^{\alpha}+{\mathbf h}_A^{\beta}+{\mathbf h}_B^{\alpha}+{\mathbf h}_B^{\beta})
\nonumber \\ &
- ({\mathbf P}^{aA} {\mathbf S}^{AO} {\mathbf P}^{iB}) \cdot ({\mathbf h}_A^{\alpha}+{\mathbf h}_B^{\alpha})
- ({\mathbf P}^{iA} {\mathbf S}^{AO} {\mathbf P}^{aB}) \cdot ({\mathbf h}_A^{\beta}+{\mathbf h}_B^{\beta})
\nonumber \\ &
+2 ({\mathbf P}^{iB} {\mathbf S}^{AO} {\mathbf P}^{iA} {\mathbf S}^{AO} {\mathbf P}^{aB}) \cdot \bm{\omega}_A
+2 ({\mathbf P}^{iB} {\mathbf S}^{AO} {\mathbf P}^{iA} {\mathbf S}^{AO} {\mathbf P}^{iB}) \cdot \bm{\omega}_A
\nonumber \\ &
+ ({\mathbf P}^{aB} {\mathbf S}^{AO} {\mathbf P}^{iA} {\mathbf S}^{AO} {\mathbf P}^{aB}) \cdot \bm{\omega}_A
+ ({\mathbf P}^{iB} {\mathbf S}^{AO} {\mathbf P}^{aA} {\mathbf S}^{AO} {\mathbf P}^{iB}) \cdot \bm{\omega}_A
\nonumber \\ &
+2 ({\mathbf P}^{iA} {\mathbf S}^{AO} {\mathbf P}^{iB} {\mathbf S}^{AO} {\mathbf P}^{iA}) \cdot \bm{\omega}_B
+2 ({\mathbf P}^{iA} {\mathbf S}^{AO} {\mathbf P}^{iB} {\mathbf S}^{AO} {\mathbf P}^{aA}) \cdot \bm{\omega}_B
\nonumber \\ &
+ ({\mathbf P}^{iA} {\mathbf S}^{AO} {\mathbf P}^{aB} {\mathbf S}^{AO} {\mathbf P}^{iA}) \cdot \bm{\omega}_B
+ ({\mathbf P}^{aA} {\mathbf S}^{AO} {\mathbf P}^{iB} {\mathbf S}^{AO} {\mathbf P}^{aA}) \cdot \bm{\omega}_B
\nonumber \\ &
-2 ({\mathbf P}^{iA} {\mathbf S}^{AO} {\mathbf P}^{iB}) \cdot {\mathbf K}[{\mathbf P}^{iA} {\mathbf S}^{AO} {\mathbf P}^{iB}]
-2 ({\mathbf P}^{aA} {\mathbf S}^{AO} {\mathbf P}^{iB}) \cdot {\mathbf K}[{\mathbf P}^{iA} {\mathbf S}^{AO} {\mathbf P}^{iB}]
\nonumber \\ &
-2 ({\mathbf P}^{iA} {\mathbf S}^{AO} {\mathbf P}^{aB}) \cdot {\mathbf K}[{\mathbf P}^{iA} {\mathbf S}^{AO} {\mathbf P}^{iB}]
- ({\mathbf P}^{aA} {\mathbf S}^{AO} {\mathbf P}^{iB}) \cdot {\mathbf K}[{\mathbf P}^{aA} {\mathbf S}^{AO} {\mathbf P}^{iB}]
\nonumber \\ &
- ({\mathbf P}^{iA} {\mathbf S}^{AO} {\mathbf P}^{aB}) \cdot {\mathbf K}[{\mathbf P}^{iA} {\mathbf S}^{AO} {\mathbf P}^{aB}]
\label{aodiag}
\end{align}
\begin{align}
E^{(10)}_{\rm exch,1flip}=&
- {\mathbf P}^{aB} \cdot {\mathbf K}^{aA}
- ({\mathbf P}^{aA} {\mathbf S}^{AO} {\mathbf P}^{aB}) \cdot ({\mathbf h}_A^{\alpha}+{\mathbf h}_B^{\beta}) 
\nonumber \\ &
+ ({\mathbf P}^{aA} {\mathbf S}^{AO} {\mathbf P}^{iB}) \cdot {\mathbf K}^{aB}
+ ({\mathbf P}^{iA} {\mathbf S}^{AO} {\mathbf P}^{aB}) \cdot {\mathbf K}^{aA}
\nonumber \\ &
+2 ({\mathbf P}^{iB} {\mathbf S}^{AO} {\mathbf P}^{aA} {\mathbf S}^{AO} {\mathbf P}^{aB}) \cdot \bm{\omega}_A
+ ({\mathbf P}^{aB} {\mathbf S}^{AO} {\mathbf P}^{aA} {\mathbf S}^{AO} {\mathbf P}^{aB}) \cdot \bm{\omega}_A
\nonumber \\ &
+2 ({\mathbf P}^{iA} {\mathbf S}^{AO} {\mathbf P}^{aB} {\mathbf S}^{AO} {\mathbf P}^{aA}) \cdot \bm{\omega}_B
+ ({\mathbf P}^{aA} {\mathbf S}^{AO} {\mathbf P}^{aB} {\mathbf S}^{AO} {\mathbf P}^{aA}) \cdot \bm{\omega_B}
\nonumber \\ &
-2 ({\mathbf P}^{aA} {\mathbf S}^{AO} {\mathbf P}^{aB}) \cdot {\mathbf K}[{\mathbf P}^{iA} {\mathbf S}^{AO} {\mathbf P}^{iB}]
-2 ({\mathbf P}^{aA} {\mathbf S}^{AO} {\mathbf P}^{aB}) \cdot {\mathbf K}[{\mathbf P}^{iA} {\mathbf S}^{AO} {\mathbf P}^{aB}]
\nonumber \\ &
-2 ({\mathbf P}^{aA} {\mathbf S}^{AO} {\mathbf P}^{iB}) \cdot {\mathbf K}[{\mathbf P}^{aA} {\mathbf S}^{AO} {\mathbf P}^{aB}]
-2 ({\mathbf P}^{aA} {\mathbf S}^{AO} {\mathbf P}^{iB}) \cdot {\mathbf K}[{\mathbf P}^{iA} {\mathbf S}^{AO} {\mathbf P}^{aB}]
\nonumber \\ &
- ({\mathbf P}^{aA} {\mathbf S}^{AO} {\mathbf P}^{aB}) \cdot {\mathbf K}[{\mathbf P}^{aA} {\mathbf S}^{AO} {\mathbf P}^{aB}]
\label{aoflip}
\end{align}
where ${\mathbf S}^{AO}$ is the overlap matrix in the AO basis.

\section{Computational Details}

The MO-based formulas for the arbitrary-spin first-order exchange energy, Eqs.~(\ref{e1exdiag}) and (\ref{offdiag1final}), 
have been implemented in two versions to verify the numerical correctness of the codes: into the ROHF-based SAPT code of
Ref.~\cite{Zuchowski:08} forming a part of the {\sc SAPT2012} package \cite{SAPT2012}, and into the development version of
the {\sc psi4} package \cite{Parrish:17} using the straightforward {\sc psi4numpy} framework \cite{Smith:18} in which 
each term corresponds to a single line of Python code. 
The exchange energy for the high-spin state of the dimer could also be computed with the code of Ref.~\cite{Zuchowski:08}
and we have verified that our new code gives identical results in the high-spin case. In fact, for two interacting
doublets, the formulas for $E^{(10)}_{\rm exch,diag}$ and $E^{(10)}_{\rm exch,1flip}$ (Eqs.~(\ref{e1exdiag}) and (\ref{offdiag1final}),
respectively) add up exactly to the formula for $E^{(10)}_{\rm exch}$ in the high-spin (triplet) state in accordance
with Eq.~(\ref{clebsch}) for $S_A=S_B=1/2$, $S=1$.
The AO-based first-order exchange expressions, 
Eqs.~(\ref{aodiag}) and (\ref{aoflip}), have been implemented into the development version of {\sc psi4}, making use of its 
efficient evaluation of generalized Coulomb and exchange matrices, both with and without the DF
approximation. Whenever the latter approximation was applied, the standard aug-cc-pV$X$Z/JKFIT set \cite{Weigend:02}
was used as the auxiliary basis accompanying the orbital set aug-cc-pV$X$Z \cite{Kendall:92}. As shown in Sec.~\ref{sec:phenalenyl}, 
the DF approximation to $E^{(10)}_{\rm exch}$ performs very well, and the density fitted AO-based code
can be applied to much larger systems than the ones studied here. 
Unless stated otherwise, we used the aug-cc-pVTZ basis set.

\section{Numerical Results}
In this section, we study the applications of theory developed in Sec. \ref{theory} to several test systems. 
For the first one, the Li$\cdots$H complex, it is possible to compare the spin-flip theory with exact SAPT calculations 
of Patkowski et al.~\cite{Patkowski:01}. Then, we focus on other atom-atom complexes including the lithium dimer, 
Li$\cdots$N, and N$\cdots$N systems. We also examine the  Mn$\cdots$Mn system,
very extensively studied in the past, for which the spin-exchange splitting is surprisingly low. Finally, we also show a test of 
exchange splittings for molecular cases: the O$_2$($^3\Sigma^-_g$) dimer and the phenalenyl dimer.

For the atomic  Li$\cdots$Li, Li$\cdots$N, and N$\cdots$N systems we can compare the energy differences between low-spin and high-spin dimers 
with results obtained from multireference {\em ab initio} methods: complete active space self-consistent field (CASSCF) and internally contracted 
multireference configuration interaction (MRCI) with the Davidson correction implemented in {\sc molpro} 2012 \cite{Werner:12}. 
The exchange splitting $\Delta E$ between the interaction energies for the highest ($E_{\rm int} (S_A+S_B) $) and lowest ($E_{\rm int} (|S_A-S_B|) $ ) multiplicities from {\em ab initio} calculations can be compared with the spin-flip SAPT term, Eq.~(\ref{offdiag1final}), 
by multiplying the latter by a following factor: 
\begin{equation}
\Delta E^{(10)}  = (1+ \frac{1}{ 2 \max{(S_A,S_B)} } ) E^{(10)}_{\rm exch,1flip},
\end{equation}
where from now on we use $\Delta E^{(10)}$ to denote the difference between the highest- and lowest multiplicity of a complex at the
SAPT level introduced in this work.


\subsection{Li$\cdots$H}

\begin{figure}[h]
\caption{First-order exchange energy for the singlet and triplet states of the Li$\cdots$H complex, computed within the
ROHF-based approach of this work (points) and the FCI-based SAPT of Ref.~\cite{Patkowski:01} (lines). The singlet-triplet
splitting from the FCI calculations is shown for comparison.}
\label{fig:lih}
\bigskip 
\begin{center}
\includegraphics[width=\textwidth]{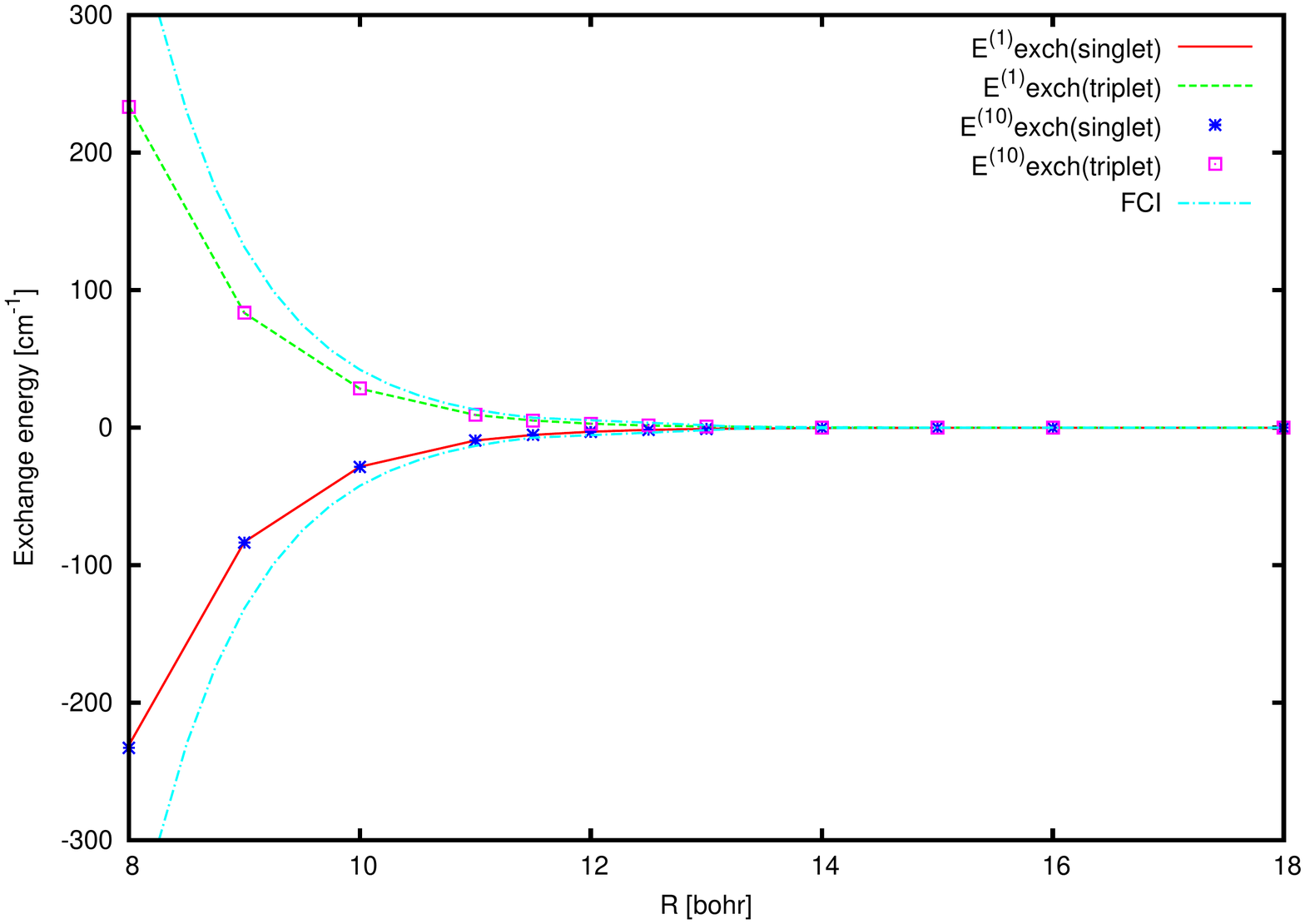}
\end{center}
\end{figure}

The first system we investigate is the Li$\cdots$H complex, where the interaction between two doublets gives rise to
a triplet state and a singlet state. For both of them, full configuration interaction (FCI) interaction energies as
well as SAPT corrections to high order (in several variants including SRS) have been obtained before \cite{Patkowski:01}.
Therefore, to facilitate our comparisons with the data of Ref.~\cite{Patkowski:01}, we use the same (4s4p1d/5s2p) basis
set in our first-order exchange calculations. The exchange energies of both spin states as functions
of the interatomic separation $R$ are presented in Fig.~\ref{fig:lih}. The excellent agreement between the results of
Ref.~\cite{Patkowski:01} (where the electron correlation within the lithium atom was described at the FCI level) and
our new values (where the Li atom was described by the ROHF theory) not only validates our implementation, but also
indicates that the intramonomer correlation effect on $E^{(1)}_{\rm exch}$ is negligible for this system. 

At the
range of distances presented in Fig.~\ref{fig:lih}, the diagonal exchange energy term, Eq.~(\ref{aodiag}), is much
smaller than the spin-flip term, Eq.~(\ref{aoflip}). The $E^{(10)}_{\rm exch,diag}$ contribution stems from 
the difference between the ``apparent Coulomb energy'', the arithmetic mean of the singlet
and triplet energies, and the actual mathematical Coulomb energy, defined in SAPT as a weighted average of the energies
of all asymptotically degenerate states contributing to the zeroth-order wavefunction, including the Pauli-forbidden 
ones \cite{Kutzelnigg:80,Patkowski:01}. It can be shown that, in the first order, the mathematical Coulomb energy is
identical to the electrostatic correction $E^{(10)}_{\rm elst}$ (Eq.~(\ref{ren1b})) up to terms that vanish as the
fourth or higher powers of intermolecular overlap integrals \cite{Jeziorski:priv}.
The apparent Coulomb
energy does not have this property and the difference between the two (that is, $E^{(10)}_{\rm exch,diag}$) 
does not vanish in the $S^2$ approximation. The only exception are the interactions involving
one- and two-electron systems, such as H$\cdots$H \cite{Cwiok:92}, He ($^3S$)$\cdots$H \cite{Przybytek:04}, and
He ($^3S$)$\cdots$He ($^3S$) \cite{Przybytek:05}, when the mathematical Coulomb energy coincides with the mean energy
of physical states. In these cases, 
$E^{(10)}_{\rm exch,diag}$ (Eq.~(\ref{e1exdiag})) is identically zero as there are no inactive orbitals 
(the range of summation over $i$ and $j$ is
empty). For the Li$\cdots$H complex, as already observed in Ref.~\cite{Patkowski:01}, 
the two Coulomb energies are particularly close and the $E^{(10)}_{\rm exch,diag}$ term is
much smaller than $E^{(10)}_{\rm exch,1flip}$.
As a result, the singlet and triplet exchange energies are almost exactly the negatives of each other. 
Assuming that this is precisely the case, one could define the exchange energy at the
FCI level as one half of the difference between the FCI interaction energies of triplet and singlet.
The FCI exchange energies defined in this way are also displayed in 
Fig.~\ref{fig:lih}. Interestingly, $E^{(10)}_{\rm exch}$ represents about two thirds of the total exchange energy,
showing that the first-order description of exchange is qualitatively correct. This result should be contrasted with
the first-order electrostatic correction $E^{(10)}_{\rm elst}$ that is nearly negligible for this dispersion-bound complex.
As a result, the sum $E^{(10)}_{\rm elst}+E^{(10)}_{\rm exch,diag}$ recovers only a small fraction 
(8\%\ at the triplet van der Waals minimum distance of 11.5$a_0$) of the FCI apparent Coulomb
energy while $E^{(10)}_{\rm exch,1flip}$ recovers 71\%\ of the FCI exchange energy.

\subsection{Li$\cdots$Li, Li$\cdots$N, and N$\cdots$N} 

\begin{figure}[h]
\caption{ Comparison of the $\Delta E^{(10)}$ energy splittings  for the Li$\cdots$Li, Li$\cdots$N, and N$\cdots$N systems with the supermolecular CASSCF and (Davidson corrected) MRCI results. For the Li dimer, experimentally derived values for the splitting given by C{\^{o}}t{\'{e}} et al. \cite{Cote:94} were also plotted.}
\label{fig:li2_lin_n2:split}
\begin{subfigure}[b]{0.55\textwidth}
\includegraphics[width=1\textwidth]{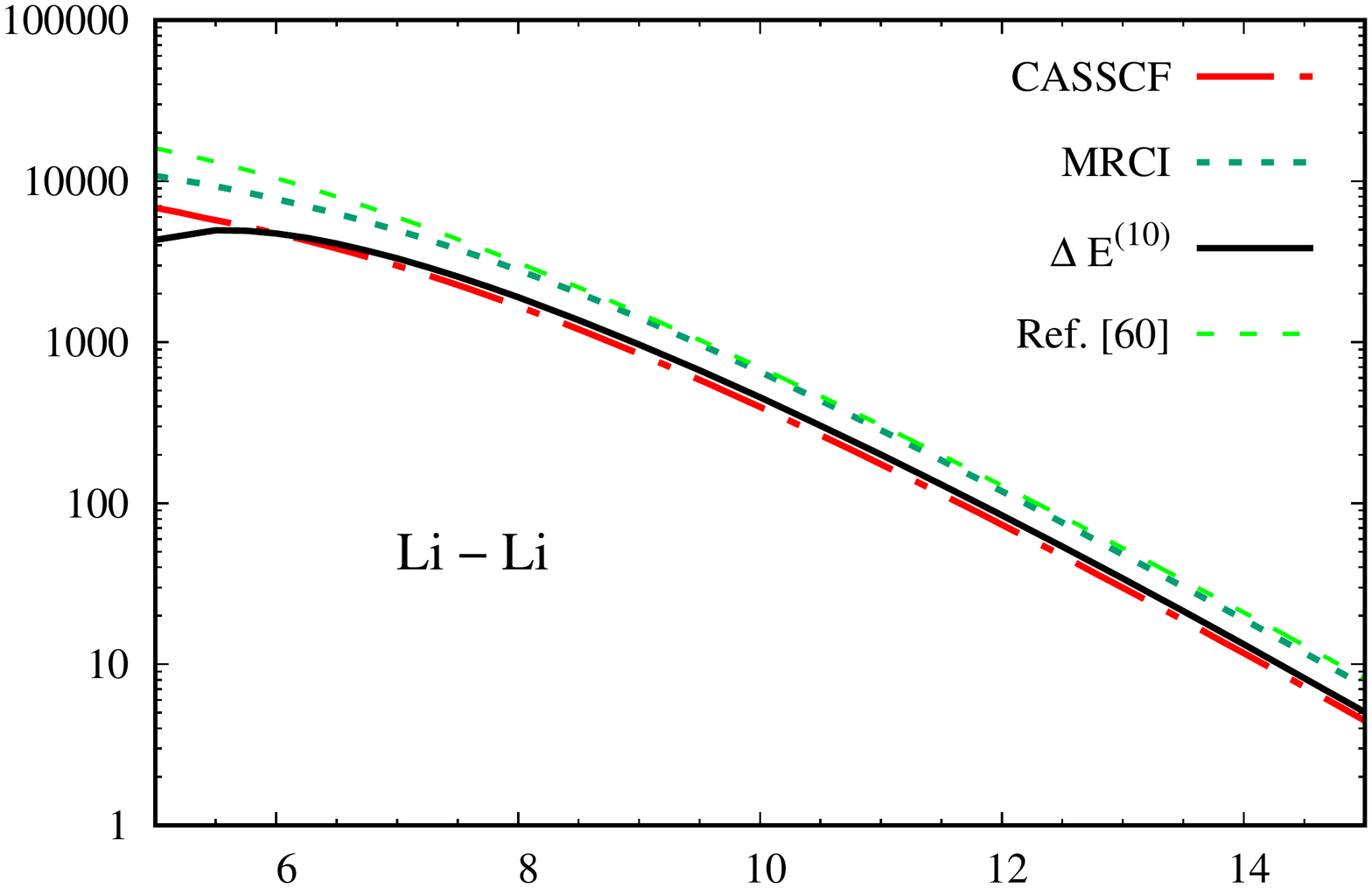}
\end{subfigure}
\begin{subfigure}[b]{0.55\textwidth}
\includegraphics[width=1\textwidth]{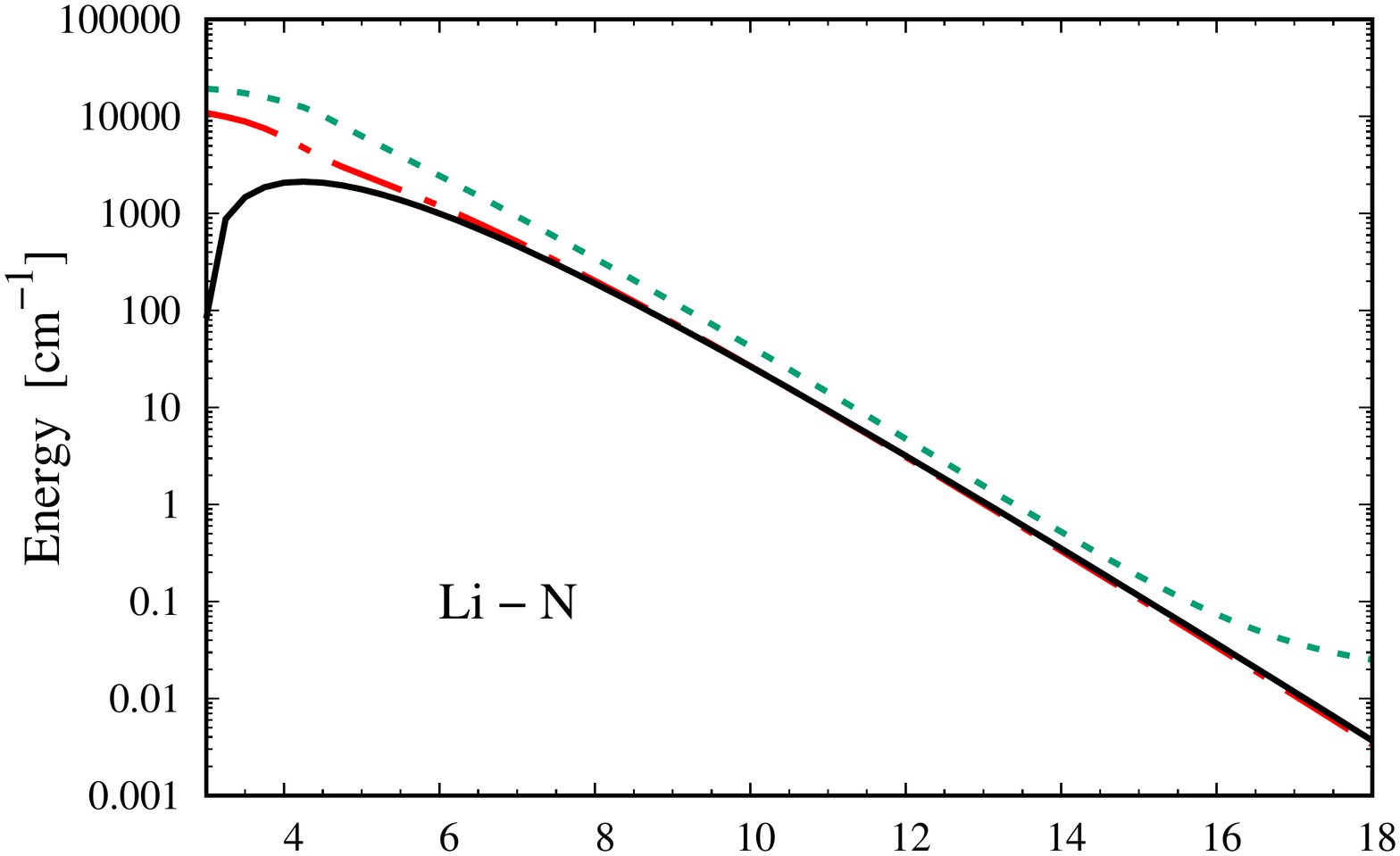}
\end{subfigure}
\begin{subfigure}[b]{0.55\textwidth}
\includegraphics[width=1\textwidth]{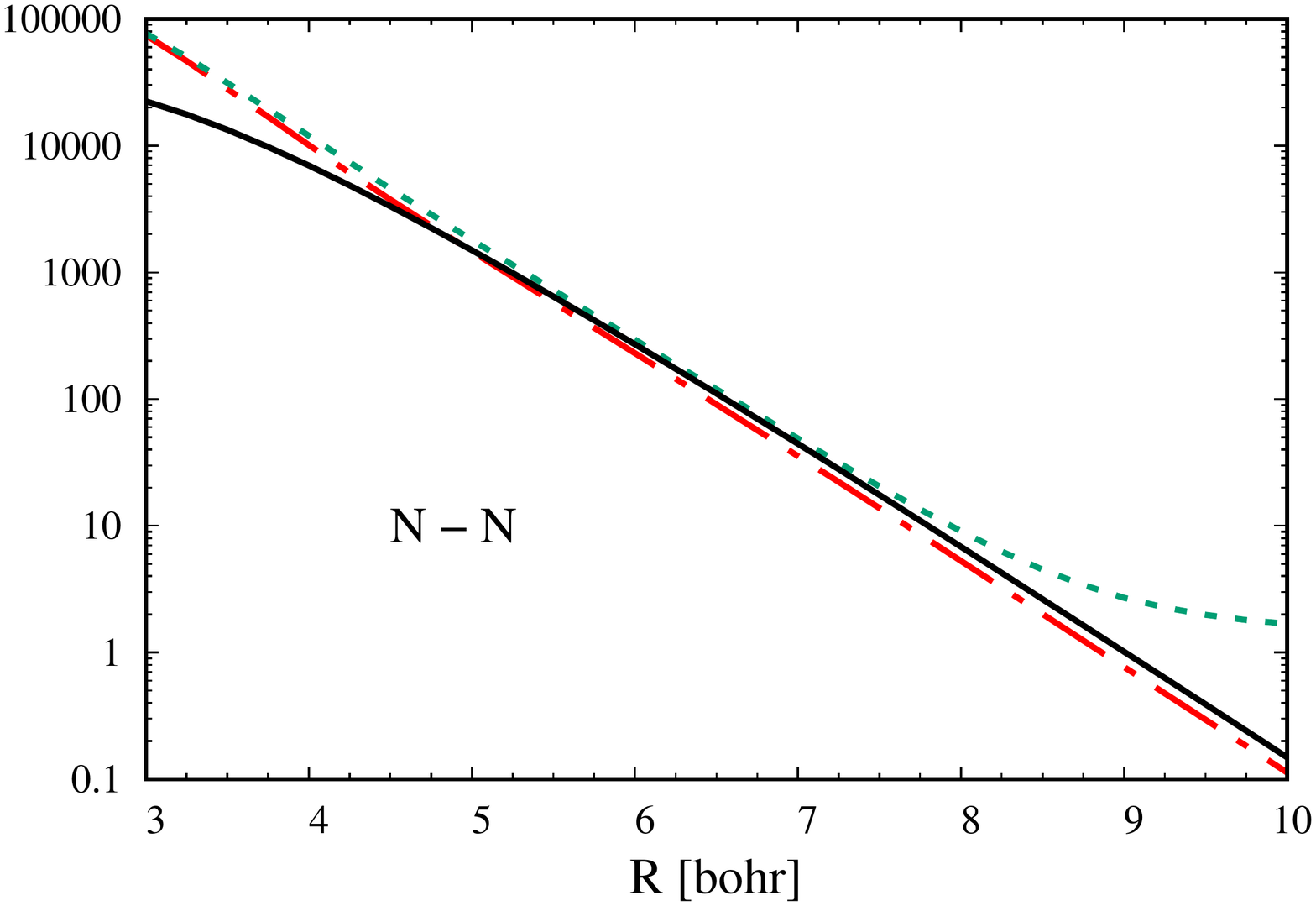}
\end{subfigure}
 \end{figure}

The total spin for the Li$\cdots$Li complex can be either singlet or triplet, for Li$\cdots$N - triplet or quintet, and for two interacting nitrogen atoms it can take multiplicities from singlet to septet. 
For these systems, unlike for the Li$\cdots$H interaction,  the benchmark SAPT corrections with exact monomer wavefunctions are not available. In Fig.~\ref{fig:li2_lin_n2:split} we show the comparison of exchange splittings from SF-SAPT, CASSCF, and MRCI (Davidson corrected).
As expected, the first-order result fails to recover the exchange splitting around the chemical minima, which are at 5$a_0$ for Li$\cdots$Li, 3.5$a_0$ for Li$\cdots$N, and 2.1$a_0$ for the nitrogen dimer. 
Nonetheless, in all these cases SF-SAPT exchange splitting is very close to the CASSCF one for the separations corresponding to the 
 van der Waals minima of the highest spin states: 7.9$a_0$ for the lithium dimer~\cite{Semczuk:13},   10.2$a_0$ for Li$\cdots$N (this value was obtained in present work using the spin-restricted CCSD(T) method with the aug-cc-pVQZ basis set), and 7.2$a_0$ for the nitrogen dimer~\cite{Tscherbul:10}.
Clearly, the exchange splitting from first-order SAPT exhibits correct asymptotic behavior. It is also worth noting that for Li$\cdots$Li,
unlike Li$\cdots$H, the diagonal and spin-flip exchange terms are of the same order as the apparent and mathematical Coulomb energies
are no longer close.
Moreover, the splittings, as well as 
$E^{(10)}_{\rm elst}$ and $E^{(10)}_{\rm exch,diag}$, show very little basis set dependence: our Li$\cdots$Li tests at $R=7.9a_0$
show that each of these three first-order contributions agrees between the (dimer-centered) cc-pVDZ and aug-cc-pVQZ basis sets to below 1\%.

In case of the lithium dimer, we could also compare the exchange splittings to the experimentally derived values given by C{\^{o}}t{\'{e}} et al. \cite{Cote:94} which confirm the correct behavior of the MRCI method. Since the $1s$ shells were frozen in MRCI
calculations, we experienced no problems with a size-inconsistent behavior of this method. Quite large differences between splittings obtained from MRCI and CASSCF/SAPT can possibly be attributed to strong influence of intramonomer dynamical correlation on the interaction effects, in particular exchange induction. 
For the Li$\cdots$N and N$\cdots$N systems, the differences between MRCI, CASSCF, and first-order SAPT are smaller. On the other hand, it is worth noting that the MRCI method  fails to reproduce the exponential decay of the exchange splitting  for large interatomic separations for Li$\cdots$N and N$\cdots$N, due to its lack of size-consistency. 

\begin{figure}[h]
\caption{ Quality of the single exchange approximation for  the high-spin first-order exchange energy in the  Li$\cdots$Li, Li$\cdots$N, and N$\cdots$N systems  measured as the ratio  $E^{(10)}_{\rm exch}(S^2)/E^{(10)}_{\rm exch}$. Color-coded vertical dashed lines correspond to the positions of minima for the high-spin complexes. }
\label{fig:li2_lin_n2:s2}
\begin{center}
\includegraphics[width=\textwidth]{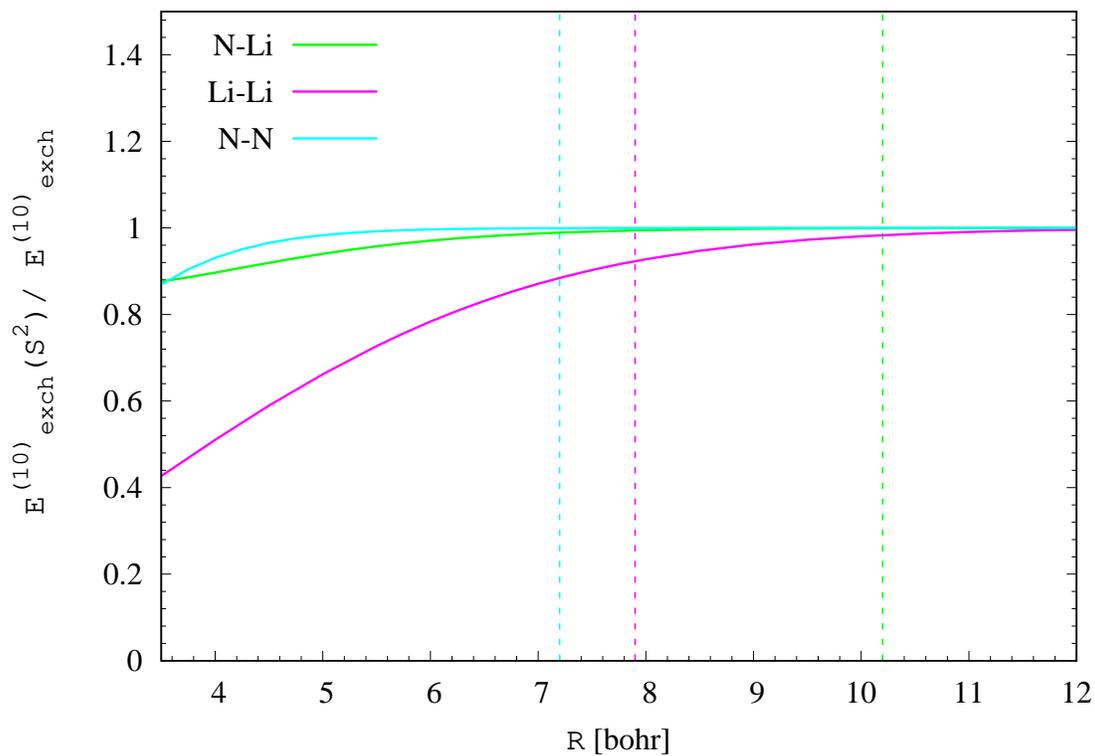}
\end{center}
\end{figure}

It might be quite useful to compare the quality of the $S^2$ approximation by inspecting the ratio of $E^{(10)}_{\rm exch}(S^2)$ and $E^{(10)}_{\rm exch}$ for the high-spin state (the nonapproximated values were computed using the high-spin code of Ref.~\cite{Zuchowski:08}). 
From Fig. \ref{fig:li2_lin_n2:s2} it is clear that while for the lithium dimer the single exchange approximation is quite drastic, even in the van der Waals minimum region, it works very well for the nitrogen dimer and the Li$\cdots$N system.  Such behavior can be attributed to a very small ionization potential $I_p$ of the Li atom (4.2 eV) compared to the N atom (14.5 eV) which directly affects the decay rates of the wavefunctions (at long range, this decay rate is proportional to $\exp{(-\sqrt{2 I_p}r)} $ \cite{Ahlrichs:81}).


\subsection{ O$_2\cdots$O$_2$}
Oxygen dimer is one of the best studied van der Waals complexes with relevance to the chemistry of atmosphere \cite{Aquilanti:99}. Hence, this was one of very first systems for which  the exchange interaction was studied. 
Wormer and van der Avoird calculated the exchange splitting within the Heisenberg model and the $J_{AB}$ parameter was obtained with variation-perturbation theory \cite{Wormer:84}. 
Later, the exchange splitting was studied by several {\em ab initio}   multireference methods.  In particular, global CASPT2 surfaces for all multiplicities were obtained by Bartolomei and coworkers \cite{Bartolomei:08}. 
To facilitate comparisons to Ref.~\cite{Bartolomei:08}, our SAPT calculations for this system use the same ANO-VTZ basis set and the same bond length in the oxygen molecules (2.28$a_0$).
 
\begin{figure}[h]
\caption{ Comparison of the first-order exchange splittings (defined here as the quintet energy minus the singlet energy) with literature data for two interacting ground-state oxygen molecules ($^3\Sigma^-_g$) for four basic geometric configurations. Since the exchange splittings for the X-shape configurations are much smaller compared to other orientations and they sometimes change sign, they are plotted on a non-logarithmic scale. The data marked `Wormer et al.' are from Ref.~\cite{Wormer:84} and the CASSCF/MRCI/CASPT2 results are from Ref.~\cite{Bartolomei:08}. }
\label{fig:O2O2fig1}
\begin{center}
\includegraphics[width=\textwidth]{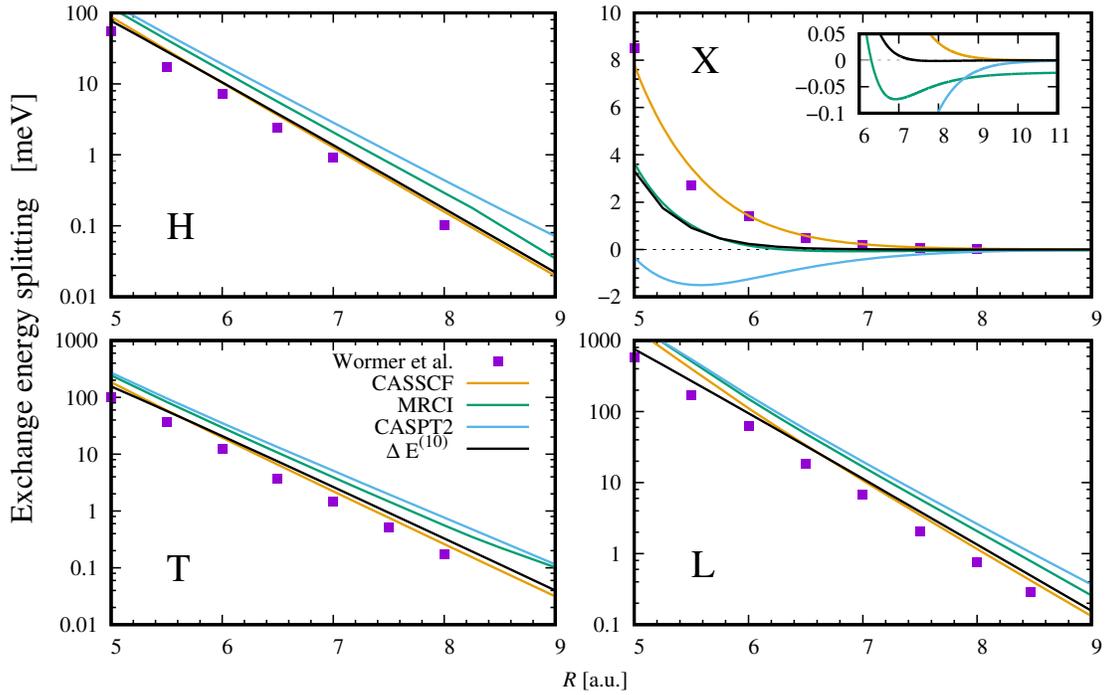}
\end{center}
\end{figure}

\begin{figure}[h]
\caption{ Diagonal and spin-flip  parts of the first-order exchange energy for the dimer of ground-state oxygen molecules ($^3\Sigma^-_g$) for four basic
geometric configurations. For the X-shape configurations, the absolute value of the spin-flip term is shown, hence the dip around 7$a_0$ corresponding to a change of sign from positive (for small $R$) to negative. }
\label{fig:O2O2fig2}
\begin{center}
\includegraphics[width=\textwidth]{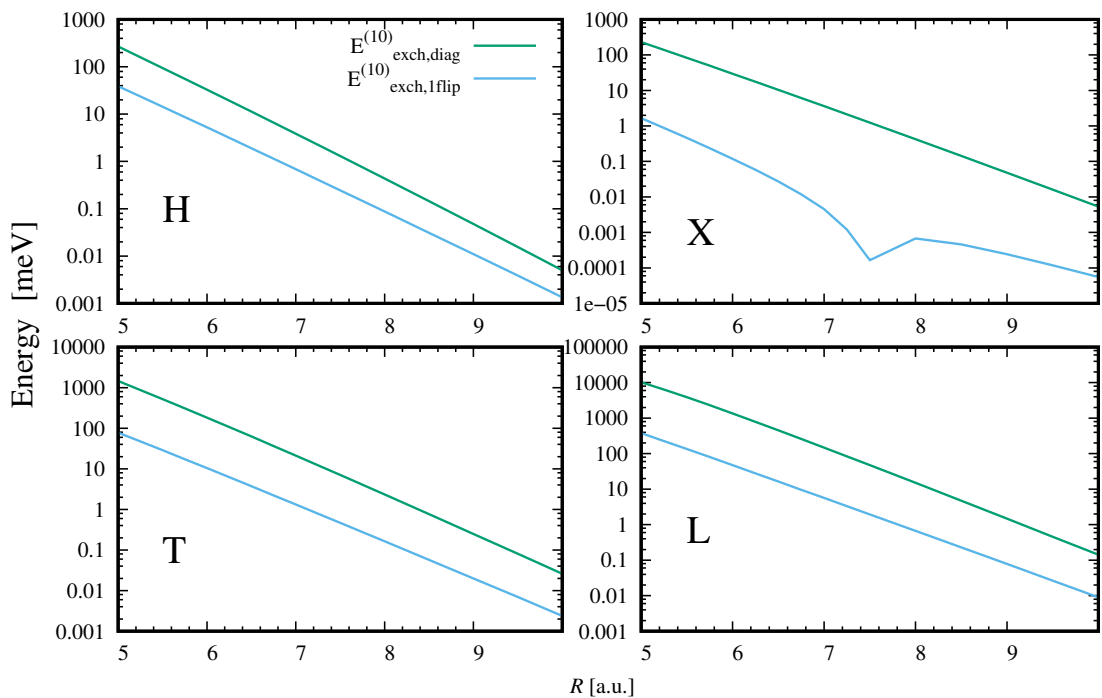}
\end{center}
\end{figure}

\begin{figure}[h]
\caption{ The quality of the $S^2$ approximation for the high-spin quintet state of the molecular oxygen dimer. The oxygen molecules are in their ground $^3\Sigma^-_g$ state and their mutual orientation corresponds to four basic
geometric configurations.}
\label{fig:O2O2fig_S2}
\begin{center}
	\includegraphics[width=\textwidth]{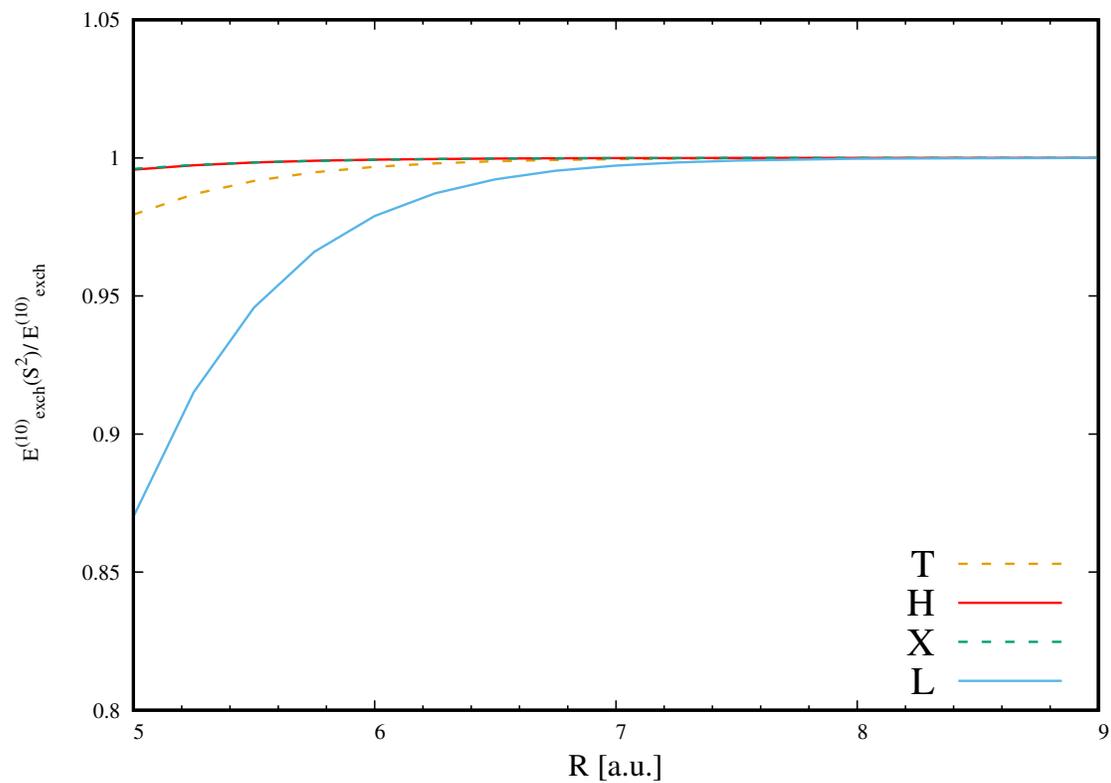}
\end{center}
\end{figure}

In Fig.~\ref{fig:O2O2fig1} we compare the splitting between the highest and lowest spin states of the O$_2\cdots$O$_2$ complex for four basic angular geometries: H-shape, linear, T-shape, and X-shape with previous studies of Wormer and van der Avoird~\cite{Wormer:84} and Bartolomei and coworkers~\cite{Bartolomei:08}, the latter including CASSCF, MRCI, and CASPT2 calculations.  
The overall agreement of the spin-flip SAPT exchange energy with these references is very satisfactory. 
The exchange splittings for the studied geometries are in a very good agreement with the Bartolomei {\em et al.} CASSCF  and Wormer and van der Avoird perturbation calculations for the H-, T-, and L-shape complexes. 
For the X-shape geometry, the first-order exchange splittings perform similarly to MRCI and are significantly  smaller compared to Heitler-London calculations of Ref.~\cite{Wormer:84} and to CASSCF. 
In the CASPT2 calculations, the exchange splittings exhibit a node for the X-shape orientation,  
which is also the case for the SAPT exchange at about 7.5$a_0$  but it is not clearly seen in Fig.~\ref{fig:O2O2fig1}. 
This node can, however, be seen on the logarithmic plot of the diagonal and spin-flip parts of the exchange splitting, Fig.~\ref{fig:O2O2fig2}, as a dip in the absolute value of the splitting. 
Note that the diagonal part of the exchange energy is very large in the oxygen dimer for all configurations including X-shape, almost two orders of magnitude larger than the spin-flip part. 
Such a big domination of the diagonal part is responsible for the fact that the oxygen molecules strongly repel each other at short range and form only weakly bound complexes for all spin states of the dimer. Finally, let us remark that the single exchange approximation works very well for the oxygen dimer, as shown by the comparison
of full and $S^2$-approximated $E^{(10)}_{\rm exch}$ values for the high-spin (quintet) complex displayed in
Fig.~\ref{fig:O2O2fig_S2}.  This behavior is consistent with the observation made by Wormer and van der Avoird~\cite{Wormer:84} that the single-parameter Heisenberg model recovers the quintet-triplet-singlet splittings accurately.

\subsection{Mn$\cdots$Mn}

The manganese dimer has attracted many studies of its exceptional exchange splitting. Its outermost electronic shell $4s$ is doubly filled and its zero orbital angular momentum ($S$ state) originates from a cancellation of half-filled $d$-shell momenta of electrons.  
For this reason, the open shells in the Mn atom are to a large extent screened by the $4s^2$ shell and the resulting spin exchange splitting is very small, on the order of 10 cm$^{-1}$ in the minimum of the potential energy curve, which is 2 orders of magnitude  less than the binding energy. 
The first {\em ab initio}  study of exchange interaction in the Mn$_2$ system was initiated by Nesbet~\cite{Nesbet:64} who used the Heisenberg model and estimated the $J_{AB}$ parameter to be small (up to 62 K at an interatomic separation of 4.5$a_0$).
More advanced methods were introduced to this system after significant progress in the multireference methods was made~\cite{Wang:04,Yamamoto:06,Negodaev:08,Tzeli:08,Camacho:08,Buchachenko:10}.  
This system is however, very challenging for multireference methods: there are 5 electrons in the submerged $d$-shell plus two electrons on the outermost $4s$ shell per atom, which makes the active space needed for dimer calculations quite large. 
Moreover, the perturbation theory approaches like CASPT2 exhibit problems for Mn$\cdots$Mn related to a presence of intruder states~\cite{Camacho:08}. 
Clearly, a development of new methods for systems similar to the manganese dimer is warranted.

\begin{figure}[h]
\caption{ First-order exchange energy for the Mn$\cdots$Mn complex: (a) diagonal and spin-flip contributions to the exchange energies compared with the 
  high-spin exchange  energy; the extremely small contribution of spin-flip exchange is evident; (b) the quality of the $S^2$ approximation for the high-spin (undecaplet) 
  Mn$\cdots$Mn  state; (c) comparison of the Heisenberg $J_{AB}$ parameters derived from SF-SAPT with existing theoretical and experimental data; note that only in the study of Buchachenko et al. \cite{Buchachenko:10} the $J_{AB}$ parameter is available as a function of $R$, whereas other data are provided for the respective equilibrium distances. For the experimental value, the distance at which $J_{AB}$ is marked corresponds to the CCSD(T) $^1\Sigma^+_g$ minimum obtained in Ref.~\cite{Buchachenko:10}. }
\label{fig:mn2}
\begin{center}
\includegraphics[width=\textwidth]{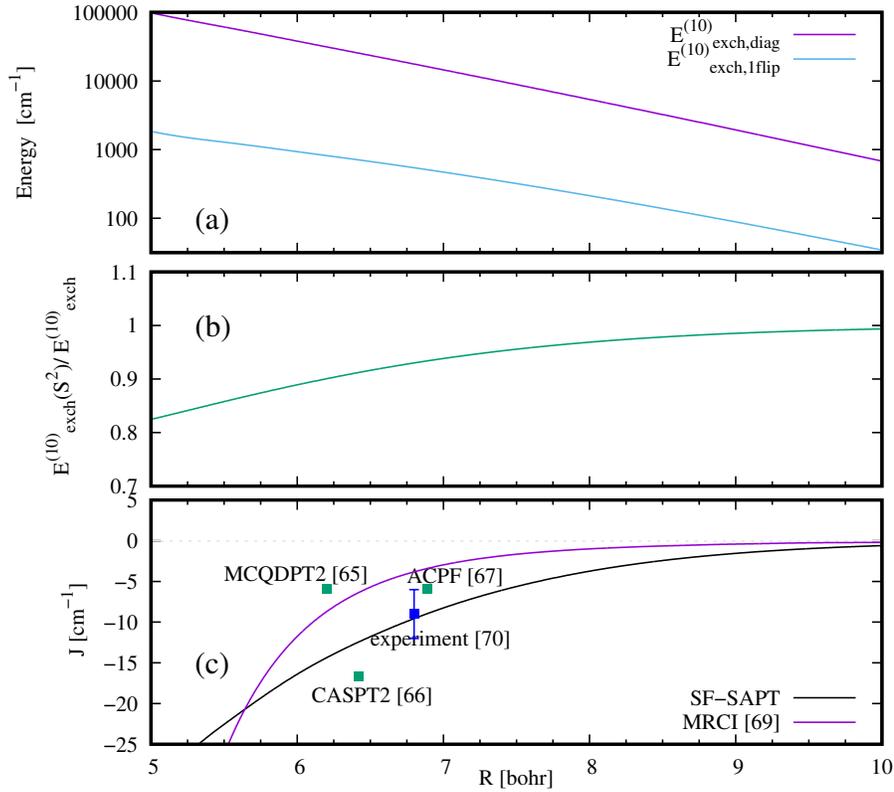}
\end{center}
\end{figure}

The ROHF method produces correct orbitals for the Mn atom (with a correct degeneracy of $d$ orbitals) and can be straightforwardly applied to first-order SF-SAPT.
In Fig.~\ref{fig:mn2}, the spin-flip SAPT exchange term is very small, nearly two orders of magnitude smaller than the diagonal exchange energy. 
Similarly to the oxygen dimer, this causes a very small exchange splitting and very small differences between the potential well depths and equilibrium distances for different multiplets. 
It is also worthwhile to inspect the quality of the $S^2$ approximation by comparing the high-spin exchange energies. Near the van der Waals minimum of undecaplet Mn$\cdots$Mn, the single exchange approximation reproduces about 90\% of the full exchange. 
In order to assess the performance of first-order SF-SAPT, we have computed the $J_{AB}$ parameter and compared it with previous literature data in panel (c) of Fig.~\ref{fig:mn2}. 
In the literature, the $J_{AB}$ parameter is usually given at the minimum separation for the high-spin complex (in case of the Buchachenko et al.~\cite{Buchachenko:10} work, the full curve was provided). 
Since the spin-flip splitting is obtained here at the ROHF level, it works remarkably well. 
In particular, our results agree very well with the experimental result of Cheeseman et al. \cite{Cheeseman:90} (although we realize this might be somewhat fortuitous). 
Except for the value by Negodaev et al. \cite{Negodaev:08}, all presented values of $J_{AB}$ are smaller than the result derived from SF-SAPT.

\subsection{Phenalenyl dimer} \label{sec:phenalenyl}

\begin{figure}[h]
\caption{The singlet-triplet energy splitting for the staggered conformation of the phenalenyl dimer. The first-order SAPT calculations and the CASSCF(2,2)
calculations use the aug-cc-pVDZ basis set. The MR-AQCC(2,2)/6-31G(d) splitting, computed in Ref.~\cite{Cui:14} for the minimum separation of the singlet
complex, is included for comparison.}
\label{fig:phen}
\begin{center}
\includegraphics[width=\textwidth]{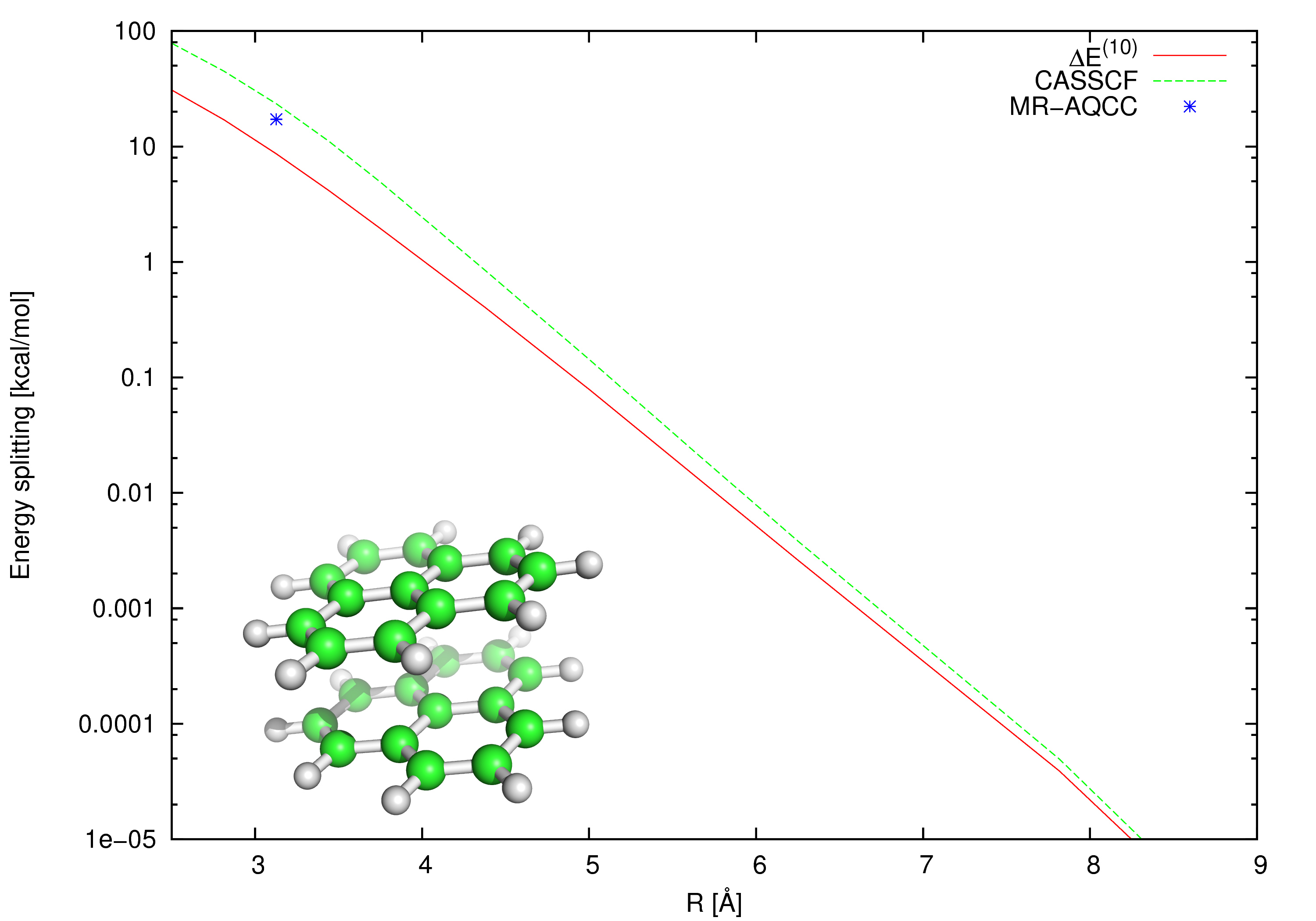}
\end{center}
\end{figure}

The dimer of the doublet phenalenyl radical is an example of ``pancake bonding'', a strong interaction between $\pi$-stacked radicals that has gathered significant interest in recent years \cite{Preuss:14}. 
The ground state of the phenalenyl dimer is a multireference singlet that exhibits pancake bonding with a binding energy of 11.5 kcal/mol, while its asymptotically degenerate triplet state exhibits only a van der Waals minimum with a depth
of 3.6 kcal/mol \cite{Cui:14}. At the interplanar separation of 3.104 \AA\ corresponding to the pancake bonded minimum depicted in Fig.~\ref{fig:phen}, the reference singlet-triplet splitting, computed using the high-level multireference
averaged quadratic coupled cluster theory (MR-AQCC) approach \cite{Szalay:93} in Ref.~\cite{Cui:14} (with a (2,2) active space and the
6-31G(d) basis set), amounts to 17.2 kcal/mol. 
It should be noted that an accurate description of pancake bonding remains a challenge to many quantum-chemical approaches, most notably those based on density functional theory \cite{Mou:17}.

In this section, we will examine how well the singlet-triplet splitting in the phenalenyl dimer is recovered by the first-order SF-SAPT approach. 
In this way, we expect to find out whether (1) the simple first-order treatment of the splitting remains valid for such a strong intermolecular attraction, (2) our AO implementation in {\sc psi4} is efficient enough to enable $E^{(10)}_{\rm exch}$ calculations for large complexes, and (3) the DF approximation is just as accurate as for conventional high-spin SAPT. For this purpose, we select the pancake-bonded minimum geometry of the singlet state as established in Ref.~\cite{Cui:14},
vary the intermolecular separation $R$, and perform MO- and AO-based SF-SAPT calculations in a number of basis sets. We did verify that, in the absence of the DF approximation, the SF-SAPT exchange energies from the MO and AO formalisms are identical. 

The singlet-triplet splittings obtained in the aug-cc-pVDZ basis set are presented in Fig.~\ref{fig:phen}. In addition to the first-order SF-SAPT calculations,
we perform CASSCF(2,2)/aug-cc-pVDZ computations across the same potential energy curve. At the pancake-bonded minimum,
the SAPT/aug-cc-pVDZ corrections $E^{(10)}_{\rm elst}$, $E^{(10)}_{\rm exch,diag}$, and
$E^{(10)}_{\rm exch,1flip}$ amount to $-19.9$, 50.6, and 4.3 kcal/mol, respectively. A basis set increase to aug-cc-pVTZ changes these values to
$-19.7$, 50.2, and 4.3 kcal/mol, respectively, confirming that the $E^{(10)}$ corrections converge quickly with the basis set size.
Thus, the singlet-triplet splitting from first-order SF-SAPT amounts to 8.6 kcal/mol or about half of the benchmark value. 
On the other hand, the CASSCF method overestimates the benchmark splitting, giving a value of 23.4 kcal/mol. Fig.~\ref{fig:phen} shows that at larger
intermonomer separations the SF-SAPT and CASSCF splittings get closer to each other and both quantities exhibit the same long-range decay.
Thus, the underestimated splitting from SF-SAPT results likely from the single-exchange approximation applied in our calculations breaking down
for the very short intermolecular distances that are characteristic of pancake bonding. 

The largest error from the DF approximation, as computed in a smaller, cc-pVDZ basis set, appears for $E^{(10)}_{\rm exch,diag}$ and amounts to 0.065 kcal/mol 
(using the cc-pVDZ/JKFIT auxiliary basis), thus, the DF approximation is fully adequate for the first-order
SF-SAPT approach. 
The cost of AO-based SF-SAPT expressions is dominated by the evaluation of generalized JK matrices, of which a total of 8 are required. As such, SF-SAPT 
takes typically about as long as 4 ROHF iterations, each of them requiring two JK builds. For the phenalenyl dimer on a six-core i7-5930K Intel CPU, 
the density-fitted evaluation of Eqs.~(\ref{aodiag})--(\ref{aoflip}) takes 26 seconds for the cc-pVDZ basis set (674 functions) and 187 seconds for the 
cc-pVTZ basis set (1556 functions), making this method tractable for very large molecules. As the algorithm uses the native JK builders of the {\sc psi4} 
code \cite{Parrish:17}, any improvements made to these core {\sc psi4} objects will also be extended to the SF-SAPT calculations.

\section{Summary}

We derived and implemented the molecular-orbital and atomic-orbital formulas for the first-order SAPT
exchange energy for two high-spin open-shell species (described by their ROHF determinants) combined to give
an arbitrary spin state of the complex (the previously existing 
open-shell SAPT approaches \cite{Zuchowski:08,Hapka:12} were restricted to the high-spin state of the complex 
except for a few system-specific studies). Within the single exchange
approximation, the resulting exchange energies for different asymptotically degenerate multiplets are
linear combinations of two common terms: the diagonal exchange (common to all spin states) and the
spin-flip term (responsible for the multiplet splittings). Thus, the single exchange approximation (which
makes double- and higher-spin-flip terms vanish identically) is equivalent to the Heisenberg Hamiltonian model
where all splittings within an asymptotically degenerate set of multiplets are expressible by a single
parameter $J_{AB}$. This equivalence was proven long ago by Matsen {\em et al.} \cite{Matsen:71,Wormer:84}, however,
this work is the first one to give an explicit, general SAPT expression for the splitting parameter.

We investigated the behavior of the diagonal and spin-flip components of the first-order SAPT exchange
correction on a number of interatomic and intermolecular complexes. 
For the Li$\cdots$H system, we compared the exchange energies with existing FCI-based SAPT 
calculations~\cite{Patkowski:01} and found a very good agreement.  In particular, ROHF-based first-order SF-SAPT 
reproduces 71\%\ of the FCI exchange splitting in the van der Waals minimum. For several other diatomic complexes:
Li$\cdots$Li, Li$\cdots$N, and N$\cdots$N we compared the difference between $E^{(10)}_{\rm exch}$ for the highest 
and lowest spin state to the splittings obtained with the CASSCF and MRCI methods. The $E^{(10)}_{\rm exch}$ 
results agree very well with CASSCF for a wide range of distances, from an infinite separation to roughly half 
the distance between the high- and  low-spin minima.  
The power of the perturbative approach is particularly 
impressive in the asymptotic region: the $E^{(10)}_{\rm exch}$ splitting very accurately reproduces 
the CASSCF values and, unlike MRCI, ensures proper exponential decay. 

For the oxygen dimer in four characteristic configurations, we compared the exchange splittings with 
literature data. Again, the first-order SAPT predictions agree very well with the MRCI and CASSCF results 
of Bartolomei et al. \cite{Bartolomei:08}. A particularly interesting test of our new theory was the manganese 
dimer, which poses a great challenge for supermolecular calculations with a variety of multireference methods. 
The Heisenberg exchange coupling constant $J_{AB}$ derived from 
$E^{(10)}_{\rm exch}$ compares qualitatively 
well with literature data and predicts small spin splitting due to screening from the outermost doubly 
occupied ($4s^2$) shell. However, it should be noted that for such a challenging system the overall agreement 
between existing supermolecular data is far from satisfactory. Interestingly, first-order spin-flip SAPT 
predicts the value of $J_{AB}$ within the experimental bounds. Finally, a highly efficient AO-based implementation 
(with density fitting) of $E^{(10)}_{\rm exch}$ allowed us to apply SF-SAPT to a larger system, the 
phenalenyl dimer. In that case, we obtained a qualitative agreement with reference CASSCF calculations.
The agreement improves at larger separations which might be attributed to the importance of terms beyond the
$S^2$ approximation. 
The new first-order exchange
corrections have been implemented into the development version of the {\sc psi4} code \cite{Parrish:17}
as well as the {\sc SAPT2012} package \cite{SAPT2012}.
The AO-based, density fitted SF-SAPT calculation is particularly efficient, leading to very fast
(and entirely single-reference) qualitatively correct estimates of the strength of multiplet splittings.

Our new development is merely the first step towards extending SAPT to arbitrary spin states of the
interacting complex. 
However, even at the present level of theory, SF-SAPT could be highly useful as a complementary method for 
transition metal complexes or potential energy surfaces near dissociation. The method introduced here can 
be also used with Kohn-Sham orbitals, provided they were obtained from a density functional which yields  
asymptotically correct exchange-correlation potentials \cite{Misquitta:05}.
The ideas presented here can be generalized to the second-order exchange-induction
and exchange-dispersion corrections as well as to a multireference, CASSCF-based description of the
noninteracting wavefunctions. Moreover, while the single exchange approximation implies that double-
and higher-spin-flip terms vanish, it is a different approximation than merely neglecting multiple
spin flips. In fact, going beyond the $S^2$ approach might be useful even for two interacting doublets
at short range (such as in the phenalenyl dimer) where there is only a single active spin on each monomer
to be flipped. The work in all of these directions is in progress in our groups.

\section*{Acknowledgments}
We are indebted to Bogumi{\l} Jeziorski for our stimulating discussions on the topic of this work.
We thank Massimiliano Bartolomei for sharing the oxygen dimer data and Alexiei Buchachenko for his comments 
regarding the Mn$\cdots$Mn dimer. K.P. and D.G.A.S. were supported by the U.S. National Science Foundation
CAREER award CHE-1351978. P.S.Z. is grateful to the Polish
National Science Center (NCN), grant number 015/19/B/ST4/02707.

\section*{Appendix: Proof of Eq.~(\ref{clebsch})}

In this Appendix, we calculate the coefficient $Z(S_A,S_B,S)=\frac{c_1}{2c_0\sqrt{S_A S_B}}$ that appears 
when inserting Eq.~(\ref{offdiag1final}) into Eq.~(\ref{e1excha}), proving Eq.~(\ref{clebsch}).
Recall that 
\begin{equation}
c_0=\langle  S\: (S_A-S_B) | S_A\: S_A\: S_B\: -S_B \rangle,
\end{equation}
\begin{equation}
c_1=\langle S\: (S_A-S_B) | S_A\: (S_A-1)\: S_B\: (-S_B+1) \rangle.
\end{equation}
In the following, we will extend our previous notation of spin raising and lowering operators 
$\hat{S}_{+}$ and $\hat{S}_{-}$, stating explicitly the molecule (A or B) that the operator is acting on. 

We start by noting that the action of the spin-flip operator $\hat{S}_{A-}\hat{S}_{B+}$ on the
initial product state $|S_A\: S_A\: S_B\: -S_B\rangle$ produces
\begin{eqnarray}
\lefteqn{\hat{S}_{A-}\hat{S}_{B+}  |S_A \: S_A\: S_B\:  -S_B \rangle}  \nonumber \\ &&
     = \sqrt{S_A(S_A+1) -S_A(S_A-1) }\sqrt{S_B(S_B+1) +S_B(1-S_B) }  |S_A\: (S_A -1)\: S_B\: (- S_B+1)  \rangle 
     \nonumber \\ && = 2\sqrt{S_AS_B}|S_A\: (S_A-1)\: S_B \: (-S_B+1) \rangle \label{eq:clebschproof1}
\end{eqnarray}

Since $\hat{S}^2=\hat{S}_A^2 + \hat{S}_B^2 + 2(\hat{S}_A)_z(\hat{S}_B)_z + \hat{S}_{A-}\hat{S}_{B+} + \hat{S}_{A+} \hat{S}_{B-}$, 
one might rewrite the action of the operator $\hat{S}_{A-}\hat{S}_{B+}$ as
\begin{eqnarray}
\lefteqn{\hat{S}_{A-}\hat{S}_{B+}  |S_A \: S_A\: S_B\:  -S_B \rangle}  \nonumber \\ && =
\left[\hat{S}^2-(\hat{S}_A^2 + \hat{S}_B^2 + 2(\hat{S}_A)_z(\hat{S}_B)_z  + \hat{S}_{A+} \hat{S}_{B-})\right] 
|S_A\: S_A\: S_B\: -S_B \rangle \nonumber \\ &&
= \left[\hat{S}^2-(S_A(S_A+1) + S_B(S_B+1) - 2S_A S_B + 0)\right] |S_A\: S_A\: S_B\: -S_B \rangle \label{eq:clebschproof2}
\end{eqnarray}
The last term in the above equation yields zero since the reverse spin-flip operator $\hat{S}_{A+}\hat{S}_{B-}$ cannot raise any more spins on A
or lower any more spins on B.
Now, by projecting Eqs.~(\ref{eq:clebschproof1}) and (\ref{eq:clebschproof2}) onto $\langle S\: (S_A-S_B) |$ and using the fact that 
$\langle S\: (S_A-S_B) | \hat{S}^2 = S(S+1)\langle S\: (S_A-S_B) |$, one finds that 
\begin{equation}
2c_1 \sqrt{S_AS_B} = \left[S(S+1) -S_A(S_A+1) - S_B(S_B+1) + 2S_A S_B\right] c_0  
\end{equation}
which, after a simple rearrangement, gives the formula (\ref{clebsch}) for $Z(S_A,S_B,S)$.

\begin{thebibliography}{77}
\expandafter\ifx\csname natexlab\endcsname\relax\def\natexlab#1{#1}\fi
\expandafter\ifx\csname bibnamefont\endcsname\relax
  \def\bibnamefont#1{#1}\fi
\expandafter\ifx\csname bibfnamefont\endcsname\relax
  \def\bibfnamefont#1{#1}\fi
\expandafter\ifx\csname citenamefont\endcsname\relax
  \def\citenamefont#1{#1}\fi
\expandafter\ifx\csname url\endcsname\relax
  \def\url#1{\texttt{#1}}\fi
\expandafter\ifx\csname urlprefix\endcsname\relax\def\urlprefix{URL }\fi
\providecommand{\bibinfo}[2]{#2}
\providecommand{\eprint}[2][]{\url{#2}}

\bibitem[{\citenamefont{Aloisio and Francisco}(2000)}]{Aloisio:00}
\bibinfo{author}{\bibfnamefont{S.}~\bibnamefont{Aloisio}} \bibnamefont{and}
  \bibinfo{author}{\bibfnamefont{J.~S.} \bibnamefont{Francisco}},
  \bibinfo{journal}{Acc. Chem. Res.} \textbf{\bibinfo{volume}{33}},
  \bibinfo{pages}{825} (\bibinfo{year}{2000}).

\bibitem[{\citenamefont{Galano et~al.}(2010)\citenamefont{Galano,
  Narciso-Lopez, and Francisco-Marquez}}]{Galano:10}
\bibinfo{author}{\bibfnamefont{A.}~\bibnamefont{Galano}},
  \bibinfo{author}{\bibfnamefont{M.}~\bibnamefont{Narciso-Lopez}},
  \bibnamefont{and}
  \bibinfo{author}{\bibfnamefont{M.}~\bibnamefont{Francisco-Marquez}},
  \bibinfo{journal}{J. Phys. Chem. A} \textbf{\bibinfo{volume}{114}},
  \bibinfo{pages}{5796} (\bibinfo{year}{2010}).

\bibitem[{\citenamefont{Roueff and Lique}(2013)}]{Roueff:13}
\bibinfo{author}{\bibfnamefont{E.}~\bibnamefont{Roueff}} \bibnamefont{and}
  \bibinfo{author}{\bibfnamefont{F.}~\bibnamefont{Lique}},
  \bibinfo{journal}{Chem. Rev.} \textbf{\bibinfo{volume}{113}},
  \bibinfo{pages}{8906} (\bibinfo{year}{2013}).

\bibitem[{\citenamefont{Ratera and Veciana}(2012)}]{Ratera:12}
\bibinfo{author}{\bibfnamefont{I.}~\bibnamefont{Ratera}} \bibnamefont{and}
  \bibinfo{author}{\bibfnamefont{J.}~\bibnamefont{Veciana}},
  \bibinfo{journal}{Chem. Soc. Rev.} \textbf{\bibinfo{volume}{41}},
  \bibinfo{pages}{303} (\bibinfo{year}{2012}).

\bibitem[{\citenamefont{Kirste et~al.}(2012)\citenamefont{Kirste, Wang, Schewe,
  Meijer, Liu, van~der Avoird, Janssen, Gubbels, Groenenboom, and van~de
  Meerakker}}]{Kirste:12}
\bibinfo{author}{\bibfnamefont{M.}~\bibnamefont{Kirste}},
  \bibinfo{author}{\bibfnamefont{X.}~\bibnamefont{Wang}},
  \bibinfo{author}{\bibfnamefont{H.~C.} \bibnamefont{Schewe}},
  \bibinfo{author}{\bibfnamefont{G.}~\bibnamefont{Meijer}},
  \bibinfo{author}{\bibfnamefont{K.}~\bibnamefont{Liu}},
  \bibinfo{author}{\bibfnamefont{A.}~\bibnamefont{van~der Avoird}},
  \bibinfo{author}{\bibfnamefont{L.~M.} \bibnamefont{Janssen}},
  \bibinfo{author}{\bibfnamefont{K.~B.} \bibnamefont{Gubbels}},
  \bibinfo{author}{\bibfnamefont{G.~C.} \bibnamefont{Groenenboom}},
  \bibnamefont{and} \bibinfo{author}{\bibfnamefont{S.~Y.} \bibnamefont{van~de
  Meerakker}}, \bibinfo{journal}{Science} \textbf{\bibinfo{volume}{338}},
  \bibinfo{pages}{1060} (\bibinfo{year}{2012}).

\bibitem[{\citenamefont{{Lavert-Ofir} et~al.}(2014)\citenamefont{{Lavert-Ofir},
  Shagam, Henson, Gersten, {K{\l}os}, {\.Zuchowski}, Narevicius, and
  Narevicius}}]{Lavert-Ofir:14}
\bibinfo{author}{\bibfnamefont{E.}~\bibnamefont{{Lavert-Ofir}}},
  \bibinfo{author}{\bibfnamefont{Y.}~\bibnamefont{Shagam}},
  \bibinfo{author}{\bibfnamefont{A.~B.} \bibnamefont{Henson}},
  \bibinfo{author}{\bibfnamefont{S.}~\bibnamefont{Gersten}},
  \bibinfo{author}{\bibfnamefont{J.}~\bibnamefont{{K{\l}os}}},
  \bibinfo{author}{\bibfnamefont{P.~S.} \bibnamefont{{\.Zuchowski}}},
  \bibinfo{author}{\bibfnamefont{J.}~\bibnamefont{Narevicius}},
  \bibnamefont{and}
  \bibinfo{author}{\bibfnamefont{E.}~\bibnamefont{Narevicius}},
  \bibinfo{journal}{Nature Chem.} \textbf{\bibinfo{volume}{6}},
  \bibinfo{pages}{332} (\bibinfo{year}{2014}).

\bibitem[{\citenamefont{Chefdeville et~al.}(2013)\citenamefont{Chefdeville,
  Kalugina, van~de Meerakker, Naulin, Lique, and Costes}}]{Chefdeville:13}
\bibinfo{author}{\bibfnamefont{S.}~\bibnamefont{Chefdeville}},
  \bibinfo{author}{\bibfnamefont{Y.}~\bibnamefont{Kalugina}},
  \bibinfo{author}{\bibfnamefont{S.~Y.~T.} \bibnamefont{van~de Meerakker}},
  \bibinfo{author}{\bibfnamefont{C.}~\bibnamefont{Naulin}},
  \bibinfo{author}{\bibfnamefont{F.}~\bibnamefont{Lique}}, \bibnamefont{and}
  \bibinfo{author}{\bibfnamefont{M.}~\bibnamefont{Costes}},
  \bibinfo{journal}{Science} \textbf{\bibinfo{volume}{341}},
  \bibinfo{pages}{1094} (\bibinfo{year}{2013}).

\bibitem[{\citenamefont{Knowles et~al.}(1993)\citenamefont{Knowles, Hampel, and
  Werner}}]{Knowles:93}
\bibinfo{author}{\bibfnamefont{P.~J.} \bibnamefont{Knowles}},
  \bibinfo{author}{\bibfnamefont{C.}~\bibnamefont{Hampel}}, \bibnamefont{and}
  \bibinfo{author}{\bibfnamefont{H.-J.} \bibnamefont{Werner}},
  \bibinfo{journal}{J. Chem. Phys.} \textbf{\bibinfo{volume}{99}},
  \bibinfo{pages}{5219} (\bibinfo{year}{1993}).

\bibitem[{\citenamefont{Watts et~al.}(1993)\citenamefont{Watts, Gauss, and
  Bartlett}}]{Watts:93}
\bibinfo{author}{\bibfnamefont{J.~D.} \bibnamefont{Watts}},
  \bibinfo{author}{\bibfnamefont{J.}~\bibnamefont{Gauss}}, \bibnamefont{and}
  \bibinfo{author}{\bibfnamefont{R.~J.} \bibnamefont{Bartlett}},
  \bibinfo{journal}{J. Chem. Phys.} \textbf{\bibinfo{volume}{98}},
  \bibinfo{pages}{8718} (\bibinfo{year}{1993}).

\bibitem[{\citenamefont{Jeziorski and Monkhorst}(1981)}]{Jeziorski:81}
\bibinfo{author}{\bibfnamefont{B.}~\bibnamefont{Jeziorski}} \bibnamefont{and}
  \bibinfo{author}{\bibfnamefont{H.~J.} \bibnamefont{Monkhorst}},
  \bibinfo{journal}{Phys. Rev. A} \textbf{\bibinfo{volume}{24}},
  \bibinfo{pages}{1668} (\bibinfo{year}{1981}).

\bibitem[{\citenamefont{Bartlett and {Musia{\l}}}(2007)}]{Bartlett:07}
\bibinfo{author}{\bibfnamefont{R.~J.} \bibnamefont{Bartlett}} \bibnamefont{and}
  \bibinfo{author}{\bibfnamefont{M.}~\bibnamefont{{Musia{\l}}}},
  \bibinfo{journal}{Rev. Mod. Phys.} \textbf{\bibinfo{volume}{79}},
  \bibinfo{pages}{291} (\bibinfo{year}{2007}).

\bibitem[{\citenamefont{Jeziorski}(2010)}]{Jeziorski:10}
\bibinfo{author}{\bibfnamefont{B.}~\bibnamefont{Jeziorski}},
  \bibinfo{journal}{Mol. Phys.} \textbf{\bibinfo{volume}{108}},
  \bibinfo{pages}{3043} (\bibinfo{year}{2010}).

\bibitem[{\citenamefont{Werner and Knowles}(1988)}]{Werner:88}
\bibinfo{author}{\bibfnamefont{H.-J.} \bibnamefont{Werner}} \bibnamefont{and}
  \bibinfo{author}{\bibfnamefont{P.~J.} \bibnamefont{Knowles}},
  \bibinfo{journal}{J. Chem. Phys.} \textbf{\bibinfo{volume}{89}},
  \bibinfo{pages}{5803} (\bibinfo{year}{1988}).

\bibitem[{\citenamefont{Andersson et~al.}(1990)\citenamefont{Andersson,
  Malmqvist, Roos, Sadlej, and Wolinski}}]{Andersson:90}
\bibinfo{author}{\bibfnamefont{K.}~\bibnamefont{Andersson}},
  \bibinfo{author}{\bibfnamefont{P.-A.} \bibnamefont{Malmqvist}},
  \bibinfo{author}{\bibfnamefont{B.}~\bibnamefont{Roos}},
  \bibinfo{author}{\bibfnamefont{A.}~\bibnamefont{Sadlej}}, \bibnamefont{and}
  \bibinfo{author}{\bibfnamefont{K.}~\bibnamefont{Wolinski}},
  \bibinfo{journal}{J. Phys. Chem.} \textbf{\bibinfo{volume}{94}},
  \bibinfo{pages}{5483} (\bibinfo{year}{1990}).

\bibitem[{\citenamefont{Finley et~al.}(1998)\citenamefont{Finley, Malmqvist,
  Roos, and {Serrano-Andr\'es}}}]{Finley:98}
\bibinfo{author}{\bibfnamefont{J.}~\bibnamefont{Finley}},
  \bibinfo{author}{\bibfnamefont{P.}~\bibnamefont{Malmqvist}},
  \bibinfo{author}{\bibfnamefont{B.~O.} \bibnamefont{Roos}}, \bibnamefont{and}
  \bibinfo{author}{\bibfnamefont{L.}~\bibnamefont{{Serrano-Andr\'es}}},
  \bibinfo{journal}{Chem. Phys. Lett.} \textbf{\bibinfo{volume}{288}},
  \bibinfo{pages}{299} (\bibinfo{year}{1998}).

\bibitem[{\citenamefont{Angeli et~al.}(2002)\citenamefont{Angeli, Cimiraglia,
  and Malrieu}}]{Angeli:02}
\bibinfo{author}{\bibfnamefont{C.}~\bibnamefont{Angeli}},
  \bibinfo{author}{\bibfnamefont{R.}~\bibnamefont{Cimiraglia}},
  \bibnamefont{and} \bibinfo{author}{\bibfnamefont{J.-P.}
  \bibnamefont{Malrieu}}, \bibinfo{journal}{J. Chem. Phys.}
  \textbf{\bibinfo{volume}{117}}, \bibinfo{pages}{9138} (\bibinfo{year}{2002}).

\bibitem[{\citenamefont{Camacho et~al.}(2010)\citenamefont{Camacho, Cimiraglia,
  and Witek}}]{Camacho:10}
\bibinfo{author}{\bibfnamefont{C.}~\bibnamefont{Camacho}},
  \bibinfo{author}{\bibfnamefont{R.}~\bibnamefont{Cimiraglia}},
  \bibnamefont{and} \bibinfo{author}{\bibfnamefont{H.~A.} \bibnamefont{Witek}},
  \bibinfo{journal}{Phys. Chem. Chem. Phys.} \textbf{\bibinfo{volume}{12}},
  \bibinfo{pages}{5058} (\bibinfo{year}{2010}).

\bibitem[{\citenamefont{Krylov}(2001)}]{Krylov:01}
\bibinfo{author}{\bibfnamefont{A.~I.} \bibnamefont{Krylov}},
  \bibinfo{journal}{Chem. Phys. Lett.} \textbf{\bibinfo{volume}{338}},
  \bibinfo{pages}{375} (\bibinfo{year}{2001}).

\bibitem[{\citenamefont{Shao et~al.}(2003)\citenamefont{Shao, {Head-Gordon},
  and Krylov}}]{Shao:03}
\bibinfo{author}{\bibfnamefont{Y.}~\bibnamefont{Shao}},
  \bibinfo{author}{\bibfnamefont{M.}~\bibnamefont{{Head-Gordon}}},
  \bibnamefont{and} \bibinfo{author}{\bibfnamefont{A.~I.}
  \bibnamefont{Krylov}}, \bibinfo{journal}{J. Chem. Phys.}
  \textbf{\bibinfo{volume}{118}}, \bibinfo{pages}{4807} (\bibinfo{year}{2003}).

\bibitem[{\citenamefont{Levchenko and Krylov}(2004)}]{Levchenko:04}
\bibinfo{author}{\bibfnamefont{S.~V.} \bibnamefont{Levchenko}}
  \bibnamefont{and} \bibinfo{author}{\bibfnamefont{A.~I.}
  \bibnamefont{Krylov}}, \bibinfo{journal}{J. Chem. Phys.}
  \textbf{\bibinfo{volume}{120}}, \bibinfo{pages}{175} (\bibinfo{year}{2004}).

\bibitem[{\citenamefont{Krylov}(2006)}]{Krylov:06}
\bibinfo{author}{\bibfnamefont{A.~I.} \bibnamefont{Krylov}},
  \bibinfo{journal}{Acc. Chem. Res.} \textbf{\bibinfo{volume}{39}},
  \bibinfo{pages}{83} (\bibinfo{year}{2006}).

\bibitem[{\citenamefont{Zimmerman et~al.}(2012)\citenamefont{Zimmerman, Bell,
  Goldey, Bell, and {Head-Gordon}}}]{Zimmerman:12}
\bibinfo{author}{\bibfnamefont{P.~M.} \bibnamefont{Zimmerman}},
  \bibinfo{author}{\bibfnamefont{F.}~\bibnamefont{Bell}},
  \bibinfo{author}{\bibfnamefont{M.}~\bibnamefont{Goldey}},
  \bibinfo{author}{\bibfnamefont{A.~T.} \bibnamefont{Bell}}, \bibnamefont{and}
  \bibinfo{author}{\bibfnamefont{M.}~\bibnamefont{{Head-Gordon}}},
  \bibinfo{journal}{J. Chem. Phys.} \textbf{\bibinfo{volume}{137}},
  \bibinfo{pages}{164110} (\bibinfo{year}{2012}).

\bibitem[{\citenamefont{Mayhall and {Head-Gordon}}(2015)}]{Mayhall:15}
\bibinfo{author}{\bibfnamefont{N.~J.} \bibnamefont{Mayhall}} \bibnamefont{and}
  \bibinfo{author}{\bibfnamefont{M.}~\bibnamefont{{Head-Gordon}}},
  \bibinfo{journal}{J. Phys. Chem. Lett.} \textbf{\bibinfo{volume}{6}},
  \bibinfo{pages}{1982} (\bibinfo{year}{2015}).

\bibitem[{\citenamefont{Jeziorski et~al.}(1994)\citenamefont{Jeziorski,
  Moszy\'nski, and Szalewicz}}]{Jeziorski:94}
\bibinfo{author}{\bibfnamefont{B.}~\bibnamefont{Jeziorski}},
  \bibinfo{author}{\bibfnamefont{R.}~\bibnamefont{Moszy\'nski}},
  \bibnamefont{and}
  \bibinfo{author}{\bibfnamefont{K.}~\bibnamefont{Szalewicz}},
  \bibinfo{journal}{Chem. Rev.} \textbf{\bibinfo{volume}{94}},
  \bibinfo{pages}{1887} (\bibinfo{year}{1994}).

\bibitem[{\citenamefont{Hohenstein and Sherrill}(2012)}]{Hohenstein:12}
\bibinfo{author}{\bibfnamefont{E.~G.} \bibnamefont{Hohenstein}}
  \bibnamefont{and} \bibinfo{author}{\bibfnamefont{C.~D.}
  \bibnamefont{Sherrill}}, \bibinfo{journal}{WIREs Comput. Mol. Sci.}
  \textbf{\bibinfo{volume}{2}}, \bibinfo{pages}{304} (\bibinfo{year}{2012}).

\bibitem[{\citenamefont{{\.Z}uchowski et~al.}(2008)\citenamefont{{\.Z}uchowski,
  Podeszwa, Moszy{\'n}ski, Jeziorski, and Szalewicz}}]{Zuchowski:08}
\bibinfo{author}{\bibfnamefont{P.~S.} \bibnamefont{{\.Z}uchowski}},
  \bibinfo{author}{\bibfnamefont{R.}~\bibnamefont{Podeszwa}},
  \bibinfo{author}{\bibfnamefont{R.}~\bibnamefont{Moszy{\'n}ski}},
  \bibinfo{author}{\bibfnamefont{B.}~\bibnamefont{Jeziorski}},
  \bibnamefont{and}
  \bibinfo{author}{\bibfnamefont{K.}~\bibnamefont{Szalewicz}},
  \bibinfo{journal}{J. Chem. Phys.} \textbf{\bibinfo{volume}{129}},
  \bibinfo{pages}{084101} (\bibinfo{year}{2008}).

\bibitem[{\citenamefont{Hapka et~al.}(2012)\citenamefont{Hapka, {\.Zuchowski},
  {Szcz\c{e}\'sniak}, and {Cha{\l}asi\'nski}}}]{Hapka:12}
\bibinfo{author}{\bibfnamefont{M.}~\bibnamefont{Hapka}},
  \bibinfo{author}{\bibfnamefont{P.~S.} \bibnamefont{{\.Zuchowski}}},
  \bibinfo{author}{\bibfnamefont{M.~M.} \bibnamefont{{Szcz\c{e}\'sniak}}},
  \bibnamefont{and}
  \bibinfo{author}{\bibfnamefont{G.}~\bibnamefont{{Cha{\l}asi\'nski}}},
  \bibinfo{journal}{J. Chem. Phys.} \textbf{\bibinfo{volume}{137}},
  \bibinfo{pages}{164104} (\bibinfo{year}{2012}).

\bibitem[{\citenamefont{Aquilanti et~al.}(1999)\citenamefont{Aquilanti,
  Ascenzi, Bartolomei, Cappeletti, Cavalli, {de Castro Vitores}, and
  Pirani}}]{Aquilanti:99}
\bibinfo{author}{\bibfnamefont{V.}~\bibnamefont{Aquilanti}},
  \bibinfo{author}{\bibfnamefont{D.}~\bibnamefont{Ascenzi}},
  \bibinfo{author}{\bibfnamefont{M.}~\bibnamefont{Bartolomei}},
  \bibinfo{author}{\bibfnamefont{D.}~\bibnamefont{Cappeletti}},
  \bibinfo{author}{\bibfnamefont{S.}~\bibnamefont{Cavalli}},
  \bibinfo{author}{\bibfnamefont{M.}~\bibnamefont{{de Castro Vitores}}},
  \bibnamefont{and} \bibinfo{author}{\bibfnamefont{F.}~\bibnamefont{Pirani}},
  \bibinfo{journal}{J. Am. Chem. Soc.} \textbf{\bibinfo{volume}{121}},
  \bibinfo{pages}{10794} (\bibinfo{year}{1999}).

\bibitem[{\citenamefont{Bartolomei et~al.}(2008)\citenamefont{Bartolomei,
  {Hern\'andez}, {Campos-Mart\'inez}, {Carmona-Novillo}, and
  {Hern\'andez-Lamoneda}}}]{Bartolomei:08}
\bibinfo{author}{\bibfnamefont{M.}~\bibnamefont{Bartolomei}},
  \bibinfo{author}{\bibfnamefont{M.~I.} \bibnamefont{{Hern\'andez}}},
  \bibinfo{author}{\bibfnamefont{J.}~\bibnamefont{{Campos-Mart\'inez}}},
  \bibinfo{author}{\bibfnamefont{E.}~\bibnamefont{{Carmona-Novillo}}},
  \bibnamefont{and}
  \bibinfo{author}{\bibfnamefont{R.}~\bibnamefont{{Hern\'andez-Lamoneda}}},
  \bibinfo{journal}{Phys. Chem. Chem. Phys.} \textbf{\bibinfo{volume}{10}},
  \bibinfo{pages}{5374} (\bibinfo{year}{2008}).

\bibitem[{\citenamefont{Bartolomei et~al.}(2010)\citenamefont{Bartolomei,
  {Carmona-Novillo}, {Hern\'andez}, {Campos-Mart\'inez}, and
  {Hern\'andez-Lamoneda}}}]{Bartolomei:10}
\bibinfo{author}{\bibfnamefont{M.}~\bibnamefont{Bartolomei}},
  \bibinfo{author}{\bibfnamefont{E.}~\bibnamefont{{Carmona-Novillo}}},
  \bibinfo{author}{\bibfnamefont{M.~I.} \bibnamefont{{Hern\'andez}}},
  \bibinfo{author}{\bibfnamefont{J.}~\bibnamefont{{Campos-Mart\'inez}}},
  \bibnamefont{and}
  \bibinfo{author}{\bibfnamefont{R.}~\bibnamefont{{Hern\'andez-Lamoneda}}},
  \bibinfo{journal}{J. Chem. Phys.} \textbf{\bibinfo{volume}{133}},
  \bibinfo{pages}{124311} (\bibinfo{year}{2010}).

\bibitem[{\citenamefont{Cui et~al.}(2014)\citenamefont{Cui, Lischka, Beneberu,
  and Kertesz}}]{Cui:14}
\bibinfo{author}{\bibfnamefont{Z.}~\bibnamefont{Cui}},
  \bibinfo{author}{\bibfnamefont{H.}~\bibnamefont{Lischka}},
  \bibinfo{author}{\bibfnamefont{H.~Z.} \bibnamefont{Beneberu}},
  \bibnamefont{and} \bibinfo{author}{\bibfnamefont{M.}~\bibnamefont{Kertesz}},
  \bibinfo{journal}{J. Am. Chem. Soc.} \textbf{\bibinfo{volume}{136}},
  \bibinfo{pages}{5539} (\bibinfo{year}{2014}).

\bibitem[{\citenamefont{Parkes et~al.}(2015)\citenamefont{Parkes, {Sava
  Gallis}, Greathouse, and Nenoff}}]{Parkes:15}
\bibinfo{author}{\bibfnamefont{M.~V.} \bibnamefont{Parkes}},
  \bibinfo{author}{\bibfnamefont{D.~F.} \bibnamefont{{Sava Gallis}}},
  \bibinfo{author}{\bibfnamefont{J.~A.} \bibnamefont{Greathouse}},
  \bibnamefont{and} \bibinfo{author}{\bibfnamefont{T.~M.}
  \bibnamefont{Nenoff}}, \bibinfo{journal}{J. Phys. Chem. C}
  \textbf{\bibinfo{volume}{119}}, \bibinfo{pages}{6556} (\bibinfo{year}{2015}).

\bibitem[{\citenamefont{Bartenstein et~al.}(2005)\citenamefont{Bartenstein,
  Altmeyer, Riedl, Geursen, Jochim, Chin, {Hecker Denschlag}, Grimm, Simoni,
  Tiesinga et~al.}}]{Bartenstein:05}
\bibinfo{author}{\bibfnamefont{M.}~\bibnamefont{Bartenstein}},
  \bibinfo{author}{\bibfnamefont{A.}~\bibnamefont{Altmeyer}},
  \bibinfo{author}{\bibfnamefont{S.}~\bibnamefont{Riedl}},
  \bibinfo{author}{\bibfnamefont{R.}~\bibnamefont{Geursen}},
  \bibinfo{author}{\bibfnamefont{S.}~\bibnamefont{Jochim}},
  \bibinfo{author}{\bibfnamefont{C.}~\bibnamefont{Chin}},
  \bibinfo{author}{\bibfnamefont{J.}~\bibnamefont{{Hecker Denschlag}}},
  \bibinfo{author}{\bibfnamefont{R.}~\bibnamefont{Grimm}},
  \bibinfo{author}{\bibfnamefont{A.}~\bibnamefont{Simoni}},
  \bibinfo{author}{\bibfnamefont{E.}~\bibnamefont{Tiesinga}},
  \bibnamefont{et~al.}, \bibinfo{journal}{Phys. Rev. Lett.}
  \textbf{\bibinfo{volume}{94}}, \bibinfo{pages}{103201}
  (\bibinfo{year}{2005}).

\bibitem[{\citenamefont{Chin et~al.}(2004)\citenamefont{Chin, Vuleti{\'{c}},
  Kerman, Chu, Tiesinga, Leo, and Williams}}]{Chin:04}
\bibinfo{author}{\bibfnamefont{C.}~\bibnamefont{Chin}},
  \bibinfo{author}{\bibfnamefont{V.}~\bibnamefont{Vuleti{\'{c}}}},
  \bibinfo{author}{\bibfnamefont{A.~J.} \bibnamefont{Kerman}},
  \bibinfo{author}{\bibfnamefont{S.}~\bibnamefont{Chu}},
  \bibinfo{author}{\bibfnamefont{E.}~\bibnamefont{Tiesinga}},
  \bibinfo{author}{\bibfnamefont{P.~J.} \bibnamefont{Leo}}, \bibnamefont{and}
  \bibinfo{author}{\bibfnamefont{C.~J.} \bibnamefont{Williams}},
  \bibinfo{journal}{Phys. Rev. A} \textbf{\bibinfo{volume}{70}},
  \bibinfo{pages}{32701} (\bibinfo{year}{2004}).

\bibitem[{\citenamefont{Williams and Chabalowski}(2001)}]{Williams:01}
\bibinfo{author}{\bibfnamefont{H.~L.} \bibnamefont{Williams}} \bibnamefont{and}
  \bibinfo{author}{\bibfnamefont{C.~F.} \bibnamefont{Chabalowski}},
  \bibinfo{journal}{J. Phys. Chem. A} \textbf{\bibinfo{volume}{105}},
  \bibinfo{pages}{646} (\bibinfo{year}{2001}).

\bibitem[{\citenamefont{Wormer and {van der Avoird}}(1984)}]{Wormer:84}
\bibinfo{author}{\bibfnamefont{P.~E.~S.} \bibnamefont{Wormer}}
  \bibnamefont{and} \bibinfo{author}{\bibfnamefont{A.}~\bibnamefont{{van der
  Avoird}}}, \bibinfo{journal}{J. Chem. Phys.} \textbf{\bibinfo{volume}{81}},
  \bibinfo{pages}{1929} (\bibinfo{year}{1984}).

\bibitem[{\citenamefont{\'Cwiok et~al.}(1992)\citenamefont{\'Cwiok, Jeziorski,
  Ko{\l}os, Moszy\'nski, and Szalewicz}}]{Cwiok:92}
\bibinfo{author}{\bibfnamefont{T.}~\bibnamefont{\'Cwiok}},
  \bibinfo{author}{\bibfnamefont{B.}~\bibnamefont{Jeziorski}},
  \bibinfo{author}{\bibfnamefont{W.}~\bibnamefont{Ko{\l}os}},
  \bibinfo{author}{\bibfnamefont{R.}~\bibnamefont{Moszy\'nski}},
  \bibnamefont{and}
  \bibinfo{author}{\bibfnamefont{K.}~\bibnamefont{Szalewicz}},
  \bibinfo{journal}{J. Chem. Phys.} \textbf{\bibinfo{volume}{97}},
  \bibinfo{pages}{7555} (\bibinfo{year}{1992}).

\bibitem[{\citenamefont{Patkowski et~al.}(2001)\citenamefont{Patkowski, Korona,
  and Jeziorski}}]{Patkowski:01}
\bibinfo{author}{\bibfnamefont{K.}~\bibnamefont{Patkowski}},
  \bibinfo{author}{\bibfnamefont{T.}~\bibnamefont{Korona}}, \bibnamefont{and}
  \bibinfo{author}{\bibfnamefont{B.}~\bibnamefont{Jeziorski}},
  \bibinfo{journal}{J. Chem. Phys.} \textbf{\bibinfo{volume}{115}},
  \bibinfo{pages}{1137} (\bibinfo{year}{2001}).

\bibitem[{\citenamefont{Patkowski et~al.}(2002)\citenamefont{Patkowski,
  Jeziorski, Korona, and Szalewicz}}]{Patkowski:02}
\bibinfo{author}{\bibfnamefont{K.}~\bibnamefont{Patkowski}},
  \bibinfo{author}{\bibfnamefont{B.}~\bibnamefont{Jeziorski}},
  \bibinfo{author}{\bibfnamefont{T.}~\bibnamefont{Korona}}, \bibnamefont{and}
  \bibinfo{author}{\bibfnamefont{K.}~\bibnamefont{Szalewicz}},
  \bibinfo{journal}{J. Chem. Phys.} \textbf{\bibinfo{volume}{117}},
  \bibinfo{pages}{5124} (\bibinfo{year}{2002}).

\bibitem[{\citenamefont{Gniewek and Jeziorski}(2014)}]{Gniewek:14}
\bibinfo{author}{\bibfnamefont{P.}~\bibnamefont{Gniewek}} \bibnamefont{and}
  \bibinfo{author}{\bibfnamefont{B.}~\bibnamefont{Jeziorski}},
  \bibinfo{journal}{Phys. Rev. A} \textbf{\bibinfo{volume}{90}},
  \bibinfo{pages}{022506} (\bibinfo{year}{2014}).

\bibitem[{\citenamefont{Gniewek and Jeziorski}(2015)}]{Gniewek:15}
\bibinfo{author}{\bibfnamefont{P.}~\bibnamefont{Gniewek}} \bibnamefont{and}
  \bibinfo{author}{\bibfnamefont{B.}~\bibnamefont{Jeziorski}},
  \bibinfo{journal}{J. Chem. Phys.} \textbf{\bibinfo{volume}{143}},
  \bibinfo{pages}{154106} (\bibinfo{year}{2015}).

\bibitem[{\citenamefont{Gniewek and Jeziorski}(2016)}]{Gniewek:16}
\bibinfo{author}{\bibfnamefont{P.}~\bibnamefont{Gniewek}} \bibnamefont{and}
  \bibinfo{author}{\bibfnamefont{B.}~\bibnamefont{Jeziorski}},
  \bibinfo{journal}{Phys. Rev. A} \textbf{\bibinfo{volume}{94}},
  \bibinfo{pages}{042708} (\bibinfo{year}{2016}).

\bibitem[{\citenamefont{Jeziorski et~al.}(1978)\citenamefont{Jeziorski,
  Cha{\l}asi\'nski, and Szalewicz}}]{Jeziorski:78}
\bibinfo{author}{\bibfnamefont{B.}~\bibnamefont{Jeziorski}},
  \bibinfo{author}{\bibfnamefont{G.}~\bibnamefont{Cha{\l}asi\'nski}},
  \bibnamefont{and}
  \bibinfo{author}{\bibfnamefont{K.}~\bibnamefont{Szalewicz}},
  \bibinfo{journal}{Int. J. Quantum Chem.} \textbf{\bibinfo{volume}{14}},
  \bibinfo{pages}{271} (\bibinfo{year}{1978}).

\bibitem[{\citenamefont{Jeziorski et~al.}(1976)\citenamefont{Jeziorski, Bulski,
  and Piela}}]{Jeziorski:76}
\bibinfo{author}{\bibfnamefont{B.}~\bibnamefont{Jeziorski}},
  \bibinfo{author}{\bibfnamefont{M.}~\bibnamefont{Bulski}}, \bibnamefont{and}
  \bibinfo{author}{\bibfnamefont{L.}~\bibnamefont{Piela}},
  \bibinfo{journal}{Int. J. Quantum Chem.} \textbf{\bibinfo{volume}{10}},
  \bibinfo{pages}{281} (\bibinfo{year}{1976}).

\bibitem[{\citenamefont{{Sch\"{a}ffer} and Jansen}(2012)}]{Schaffer:12}
\bibinfo{author}{\bibfnamefont{R.}~\bibnamefont{{Sch\"{a}ffer}}}
  \bibnamefont{and} \bibinfo{author}{\bibfnamefont{G.}~\bibnamefont{Jansen}},
  \bibinfo{journal}{Theor. Chem. Acc.} \textbf{\bibinfo{volume}{131}},
  \bibinfo{pages}{1235} (\bibinfo{year}{2012}).

\bibitem[{\citenamefont{{Sch\"{a}ffer} and Jansen}(2013)}]{Schaffer:13}
\bibinfo{author}{\bibfnamefont{R.}~\bibnamefont{{Sch\"{a}ffer}}}
  \bibnamefont{and} \bibinfo{author}{\bibfnamefont{G.}~\bibnamefont{Jansen}},
  \bibinfo{journal}{Mol. Phys.} \textbf{\bibinfo{volume}{111}},
  \bibinfo{pages}{2570} (\bibinfo{year}{2013}).

\bibitem[{\citenamefont{Moszy\'nski et~al.}(1994)\citenamefont{Moszy\'nski,
  Jeziorski, Rybak, Szalewicz, and Williams}}]{Moszynski:94a}
\bibinfo{author}{\bibfnamefont{R.}~\bibnamefont{Moszy\'nski}},
  \bibinfo{author}{\bibfnamefont{B.}~\bibnamefont{Jeziorski}},
  \bibinfo{author}{\bibfnamefont{S.}~\bibnamefont{Rybak}},
  \bibinfo{author}{\bibfnamefont{K.}~\bibnamefont{Szalewicz}},
  \bibnamefont{and} \bibinfo{author}{\bibfnamefont{H.~L.}
  \bibnamefont{Williams}}, \bibinfo{journal}{J. Chem. Phys.}
  \textbf{\bibinfo{volume}{100}}, \bibinfo{pages}{5080} (\bibinfo{year}{1994}).

\bibitem[{\citenamefont{Patkowski et~al.}(2006)\citenamefont{Patkowski,
  Szalewicz, and Jeziorski}}]{Patkowski:06}
\bibinfo{author}{\bibfnamefont{K.}~\bibnamefont{Patkowski}},
  \bibinfo{author}{\bibfnamefont{K.}~\bibnamefont{Szalewicz}},
  \bibnamefont{and}
  \bibinfo{author}{\bibfnamefont{B.}~\bibnamefont{Jeziorski}},
  \bibinfo{journal}{J. Chem. Phys.} \textbf{\bibinfo{volume}{125}},
  \bibinfo{pages}{154107} (\bibinfo{year}{2006}).

\bibitem[{\citenamefont{Matsen et~al.}(1971)\citenamefont{Matsen, Klein, and
  Foyt}}]{Matsen:71}
\bibinfo{author}{\bibfnamefont{F.~A.} \bibnamefont{Matsen}},
  \bibinfo{author}{\bibfnamefont{D.~J.} \bibnamefont{Klein}}, \bibnamefont{and}
  \bibinfo{author}{\bibfnamefont{D.~C.} \bibnamefont{Foyt}},
  \bibinfo{journal}{J. Phys. Chem.} \textbf{\bibinfo{volume}{75}},
  \bibinfo{pages}{1866} (\bibinfo{year}{1971}).

\bibitem[{\citenamefont{Hesselmann et~al.}(2005)\citenamefont{Hesselmann,
  Jansen, and Sch{\"u}tz}}]{Hesselmann:05}
\bibinfo{author}{\bibfnamefont{A.}~\bibnamefont{Hesselmann}},
  \bibinfo{author}{\bibfnamefont{G.}~\bibnamefont{Jansen}}, \bibnamefont{and}
  \bibinfo{author}{\bibfnamefont{M.}~\bibnamefont{Sch{\"u}tz}},
  \bibinfo{journal}{J. Chem. Phys.} \textbf{\bibinfo{volume}{122}},
  \bibinfo{pages}{014103} (\bibinfo{year}{2005}).

\bibitem[{\citenamefont{Parrish et~al.}(2017)\citenamefont{Parrish, Burns,
  Smith, Simmonett, {DePrince, III}, Hohenstein, Bozkaya, Sokolov, {Di
  Remigio}, Richard et~al.}}]{Parrish:17}
\bibinfo{author}{\bibfnamefont{R.~M.} \bibnamefont{Parrish}},
  \bibinfo{author}{\bibfnamefont{L.~A.} \bibnamefont{Burns}},
  \bibinfo{author}{\bibfnamefont{D.~G.~A.} \bibnamefont{Smith}},
  \bibinfo{author}{\bibfnamefont{A.~C.} \bibnamefont{Simmonett}},
  \bibinfo{author}{\bibfnamefont{A.~E.} \bibnamefont{{DePrince, III}}},
  \bibinfo{author}{\bibfnamefont{E.~G.} \bibnamefont{Hohenstein}},
  \bibinfo{author}{\bibfnamefont{U.}~\bibnamefont{Bozkaya}},
  \bibinfo{author}{\bibfnamefont{A.~Y.} \bibnamefont{Sokolov}},
  \bibinfo{author}{\bibfnamefont{R.}~\bibnamefont{{Di Remigio}}},
  \bibinfo{author}{\bibfnamefont{R.~M.} \bibnamefont{Richard}},
  \bibnamefont{et~al.}, \bibinfo{journal}{J. Chem. Theory Comput.}
  \textbf{\bibinfo{volume}{13}}, \bibinfo{pages}{3185} (\bibinfo{year}{2017}).

\bibitem[{\citenamefont{Williams et~al.}(1995)\citenamefont{Williams, Mas,
  Szalewicz, and Jeziorski}}]{Williams:95}
\bibinfo{author}{\bibfnamefont{H.~L.} \bibnamefont{Williams}},
  \bibinfo{author}{\bibfnamefont{E.~M.} \bibnamefont{Mas}},
  \bibinfo{author}{\bibfnamefont{K.}~\bibnamefont{Szalewicz}},
  \bibnamefont{and}
  \bibinfo{author}{\bibfnamefont{B.}~\bibnamefont{Jeziorski}},
  \bibinfo{journal}{J. Chem. Phys.} \textbf{\bibinfo{volume}{103}},
  \bibinfo{pages}{7374} (\bibinfo{year}{1995}).

\bibitem[{SAP()}]{SAPT2012}
\bibinfo{note}{{\em SAPT2012: An Ab Initio Program for Many-Body
  Symmetry-Adapted Perturbation Theory Calculations of Intermolecular
  Interaction Energies,} {by R. Bukowski, W. Cencek, P. Jankowski, M.
  Jeziorska, B. Jeziorski, S. A. Kucharski, V.~F. Lotrich, A.~J. Misquitta, R.
  {Moszy\'{n}ski}, K. Patkowski, R. Podeszwa, F. Rob, S. Rybak, K. Szalewicz,
  H.~L. Williams, R.~J. Wheatley, P.~E.~S. Wormer, and P. S. {\.Z}uchowski,
  University of Delaware and University of Warsaw}
  ({http://www.physics.udel.edu/$\sim$szalewic/SAPT/SAPT.html}).}

\bibitem[{\citenamefont{Smith et~al.}(2018)\citenamefont{Smith, Burns,
  Sirianni, Nascimento, Kumar, James, Schriber, Zhang, Zhang, Abbott
  et~al.}}]{Smith:18}
\bibinfo{author}{\bibfnamefont{D.~G.~A.} \bibnamefont{Smith}},
  \bibinfo{author}{\bibfnamefont{L.~A.} \bibnamefont{Burns}},
  \bibinfo{author}{\bibfnamefont{D.~A.} \bibnamefont{Sirianni}},
  \bibinfo{author}{\bibfnamefont{D.~R.} \bibnamefont{Nascimento}},
  \bibinfo{author}{\bibfnamefont{A.}~\bibnamefont{Kumar}},
  \bibinfo{author}{\bibfnamefont{A.~M.} \bibnamefont{James}},
  \bibinfo{author}{\bibfnamefont{J.~B.} \bibnamefont{Schriber}},
  \bibinfo{author}{\bibfnamefont{T.}~\bibnamefont{Zhang}},
  \bibinfo{author}{\bibfnamefont{B.}~\bibnamefont{Zhang}},
  \bibinfo{author}{\bibfnamefont{A.~S.} \bibnamefont{Abbott}},
  \bibnamefont{et~al.} (\bibinfo{year}{2018}),
  \urlprefix\url{https://chemrxiv.org/articles/Psi4NumPy_An_Interactive_Quantum_Chemistry_Programming_Environment_for_Reference_Implementations_and_Rapid_Development/5746059}.

\bibitem[{\citenamefont{Weigend et~al.}(2002)\citenamefont{Weigend, {K\"{o}hn},
  and {H\"{a}ttig}}}]{Weigend:02}
\bibinfo{author}{\bibfnamefont{F.}~\bibnamefont{Weigend}},
  \bibinfo{author}{\bibfnamefont{A.}~\bibnamefont{{K\"{o}hn}}},
  \bibnamefont{and}
  \bibinfo{author}{\bibfnamefont{C.}~\bibnamefont{{H\"{a}ttig}}},
  \bibinfo{journal}{J. Chem. Phys.} \textbf{\bibinfo{volume}{116}},
  \bibinfo{pages}{3175} (\bibinfo{year}{2002}).

\bibitem[{\citenamefont{Kendall et~al.}(1992)\citenamefont{Kendall, {Dunning
  Jr.}, and Harrison}}]{Kendall:92}
\bibinfo{author}{\bibfnamefont{R.~A.} \bibnamefont{Kendall}},
  \bibinfo{author}{\bibfnamefont{T.~H.} \bibnamefont{{Dunning Jr.}}},
  \bibnamefont{and} \bibinfo{author}{\bibfnamefont{R.~J.}
  \bibnamefont{Harrison}}, \bibinfo{journal}{J. Chem. Phys.}
  \textbf{\bibinfo{volume}{96}}, \bibinfo{pages}{6796} (\bibinfo{year}{1992}).

\bibitem[{\citenamefont{Werner et~al.}(2012)\citenamefont{Werner, Knowles,
  Knizia, Manby, and Sch{\"u}tz}}]{Werner:12}
\bibinfo{author}{\bibfnamefont{H.-J.} \bibnamefont{Werner}},
  \bibinfo{author}{\bibfnamefont{P.~J.} \bibnamefont{Knowles}},
  \bibinfo{author}{\bibfnamefont{G.}~\bibnamefont{Knizia}},
  \bibinfo{author}{\bibfnamefont{F.~R.} \bibnamefont{Manby}}, \bibnamefont{and}
  \bibinfo{author}{\bibfnamefont{M.}~\bibnamefont{Sch{\"u}tz}},
  \bibinfo{journal}{WIREs Comput Mol Sci} \textbf{\bibinfo{volume}{2}},
  \bibinfo{pages}{242} (\bibinfo{year}{2012}).

\bibitem[{\citenamefont{Kutzelnigg}(1980)}]{Kutzelnigg:80}
\bibinfo{author}{\bibfnamefont{W.}~\bibnamefont{Kutzelnigg}},
  \bibinfo{journal}{J. Chem. Phys.} \textbf{\bibinfo{volume}{73}},
  \bibinfo{pages}{343} (\bibinfo{year}{1980}).

\bibitem[{\citenamefont{Jeziorski}()}]{Jeziorski:priv}
\bibinfo{author}{\bibfnamefont{B.}~\bibnamefont{Jeziorski}},
  \bibinfo{note}{private communication}.

\bibitem[{\citenamefont{Przybytek et~al.}(2004)\citenamefont{Przybytek,
  Patkowski, and Jeziorski}}]{Przybytek:04}
\bibinfo{author}{\bibfnamefont{M.}~\bibnamefont{Przybytek}},
  \bibinfo{author}{\bibfnamefont{K.}~\bibnamefont{Patkowski}},
  \bibnamefont{and}
  \bibinfo{author}{\bibfnamefont{B.}~\bibnamefont{Jeziorski}},
  \bibinfo{journal}{Collect. Czech. Chem. Commun.}
  \textbf{\bibinfo{volume}{69}}, \bibinfo{pages}{141} (\bibinfo{year}{2004}).

\bibitem[{\citenamefont{Przybytek and Jeziorski}(2005)}]{Przybytek:05}
\bibinfo{author}{\bibfnamefont{M.}~\bibnamefont{Przybytek}} \bibnamefont{and}
  \bibinfo{author}{\bibfnamefont{B.}~\bibnamefont{Jeziorski}},
  \bibinfo{journal}{J. Chem. Phys.} \textbf{\bibinfo{volume}{123}},
  \bibinfo{pages}{134315} (\bibinfo{year}{2005}).

\bibitem[{\citenamefont{Semczuk et~al.}(2013)\citenamefont{Semczuk, Li, Gunton,
  Haw, Dattani, Witz, Mills, Jones, and Madison}}]{Semczuk:13}
\bibinfo{author}{\bibfnamefont{M.}~\bibnamefont{Semczuk}},
  \bibinfo{author}{\bibfnamefont{X.}~\bibnamefont{Li}},
  \bibinfo{author}{\bibfnamefont{W.}~\bibnamefont{Gunton}},
  \bibinfo{author}{\bibfnamefont{M.}~\bibnamefont{Haw}},
  \bibinfo{author}{\bibfnamefont{N.~S.} \bibnamefont{Dattani}},
  \bibinfo{author}{\bibfnamefont{J.}~\bibnamefont{Witz}},
  \bibinfo{author}{\bibfnamefont{A.~K.} \bibnamefont{Mills}},
  \bibinfo{author}{\bibfnamefont{D.~J.} \bibnamefont{Jones}}, \bibnamefont{and}
  \bibinfo{author}{\bibfnamefont{K.~W.} \bibnamefont{Madison}},
  \bibinfo{journal}{Phys. Rev. A} \textbf{\bibinfo{volume}{87}},
  \bibinfo{pages}{052505} (\bibinfo{year}{2013}).

\bibitem[{\citenamefont{Tscherbul et~al.}(2010)\citenamefont{Tscherbul,
  K{\l}os, Dalgarno, Zygelman, Pavlovic, Hummon, Lu, Tsikata, and
  Doyle}}]{Tscherbul:10}
\bibinfo{author}{\bibfnamefont{T.~V.} \bibnamefont{Tscherbul}},
  \bibinfo{author}{\bibfnamefont{J.}~\bibnamefont{K{\l}os}},
  \bibinfo{author}{\bibfnamefont{A.}~\bibnamefont{Dalgarno}},
  \bibinfo{author}{\bibfnamefont{B.}~\bibnamefont{Zygelman}},
  \bibinfo{author}{\bibfnamefont{Z.}~\bibnamefont{Pavlovic}},
  \bibinfo{author}{\bibfnamefont{M.~T.} \bibnamefont{Hummon}},
  \bibinfo{author}{\bibfnamefont{H.-I.} \bibnamefont{Lu}},
  \bibinfo{author}{\bibfnamefont{E.}~\bibnamefont{Tsikata}}, \bibnamefont{and}
  \bibinfo{author}{\bibfnamefont{J.~M.} \bibnamefont{Doyle}},
  \bibinfo{journal}{Phys. Rev. A} \textbf{\bibinfo{volume}{82}},
  \bibinfo{pages}{042718} (\bibinfo{year}{2010}).

\bibitem[{\citenamefont{C{\^{o}}t{\'{e}}
  et~al.}(1994)\citenamefont{C{\^{o}}t{\'{e}}, Dalgarno, and
  Jamieson}}]{Cote:94}
\bibinfo{author}{\bibfnamefont{R.}~\bibnamefont{C{\^{o}}t{\'{e}}}},
  \bibinfo{author}{\bibfnamefont{A.}~\bibnamefont{Dalgarno}}, \bibnamefont{and}
  \bibinfo{author}{\bibfnamefont{M.~J.} \bibnamefont{Jamieson}},
  \bibinfo{journal}{Phys. Rev. A} \textbf{\bibinfo{volume}{50}},
  \bibinfo{pages}{399} (\bibinfo{year}{1994}).

\bibitem[{\citenamefont{Ahlrichs et~al.}(1981)\citenamefont{Ahlrichs,
  Hoffmann-Ostenhof, Hoffmann-Ostenhof, and Morgan}}]{Ahlrichs:81}
\bibinfo{author}{\bibfnamefont{R.}~\bibnamefont{Ahlrichs}},
  \bibinfo{author}{\bibfnamefont{M.}~\bibnamefont{Hoffmann-Ostenhof}},
  \bibinfo{author}{\bibfnamefont{T.}~\bibnamefont{Hoffmann-Ostenhof}},
  \bibnamefont{and} \bibinfo{author}{\bibfnamefont{J.~D.}
  \bibnamefont{Morgan}}, \bibinfo{journal}{Phys. Rev. A}
  \textbf{\bibinfo{volume}{23}}, \bibinfo{pages}{2106} (\bibinfo{year}{1981}).

\bibitem[{\citenamefont{Nesbet}(1964)}]{Nesbet:64}
\bibinfo{author}{\bibfnamefont{R.~K.} \bibnamefont{Nesbet}},
  \bibinfo{journal}{Phys. Rev.} \textbf{\bibinfo{volume}{135}}
  (\bibinfo{year}{1964}).

\bibitem[{\citenamefont{Wang et~al.}(2004)\citenamefont{Wang, Lucchese, and
  Bevan}}]{Wang:04}
\bibinfo{author}{\bibfnamefont{Z.}~\bibnamefont{Wang}},
  \bibinfo{author}{\bibfnamefont{R.~R.} \bibnamefont{Lucchese}},
  \bibnamefont{and} \bibinfo{author}{\bibfnamefont{J.~W.} \bibnamefont{Bevan}},
  \bibinfo{journal}{J. Phys. Chem. A} \textbf{\bibinfo{volume}{108}},
  \bibinfo{pages}{2884} (\bibinfo{year}{2004}).

\bibitem[{\citenamefont{Yamamoto et~al.}(2006)\citenamefont{Yamamoto, Tatewaki,
  Moriyama, and Nakano}}]{Yamamoto:06}
\bibinfo{author}{\bibfnamefont{S.}~\bibnamefont{Yamamoto}},
  \bibinfo{author}{\bibfnamefont{H.}~\bibnamefont{Tatewaki}},
  \bibinfo{author}{\bibfnamefont{H.}~\bibnamefont{Moriyama}}, \bibnamefont{and}
  \bibinfo{author}{\bibfnamefont{H.}~\bibnamefont{Nakano}},
  \bibinfo{journal}{J. Chem. Phys.} \textbf{\bibinfo{volume}{124}},
  \bibinfo{pages}{124302} (\bibinfo{year}{2006}).

\bibitem[{\citenamefont{Negodaev et~al.}(2008)\citenamefont{Negodaev, de~Graaf,
  and Caballol}}]{Negodaev:08}
\bibinfo{author}{\bibfnamefont{I.}~\bibnamefont{Negodaev}},
  \bibinfo{author}{\bibfnamefont{C.}~\bibnamefont{de~Graaf}}, \bibnamefont{and}
  \bibinfo{author}{\bibfnamefont{R.}~\bibnamefont{Caballol}},
  \bibinfo{journal}{Chem. Phys. Lett.} \textbf{\bibinfo{volume}{458}},
  \bibinfo{pages}{290} (\bibinfo{year}{2008}).

\bibitem[{\citenamefont{Tzeli et~al.}(2008)\citenamefont{Tzeli, Miranda,
  Kaplan, and Mavridis}}]{Tzeli:08}
\bibinfo{author}{\bibfnamefont{D.}~\bibnamefont{Tzeli}},
  \bibinfo{author}{\bibfnamefont{U.}~\bibnamefont{Miranda}},
  \bibinfo{author}{\bibfnamefont{I.~G.} \bibnamefont{Kaplan}},
  \bibnamefont{and} \bibinfo{author}{\bibfnamefont{A.}~\bibnamefont{Mavridis}},
  \bibinfo{journal}{J. Chem. Phys.} \textbf{\bibinfo{volume}{129}},
  \bibinfo{pages}{154310} (\bibinfo{year}{2008}).

\bibitem[{\citenamefont{Camacho et~al.}(2008)\citenamefont{Camacho, Yamamoto,
  and Witek}}]{Camacho:08}
\bibinfo{author}{\bibfnamefont{C.}~\bibnamefont{Camacho}},
  \bibinfo{author}{\bibfnamefont{S.}~\bibnamefont{Yamamoto}}, \bibnamefont{and}
  \bibinfo{author}{\bibfnamefont{H.~A.} \bibnamefont{Witek}},
  \bibinfo{journal}{Phys. Chem. Chem. Phys.} \textbf{\bibinfo{volume}{10}},
  \bibinfo{pages}{5128} (\bibinfo{year}{2008}).

\bibitem[{\citenamefont{Buchachenko et~al.}(2010)\citenamefont{Buchachenko,
  {Cha{\l}asi\'nski}, and {Szcz\c{e}\'sniak}}}]{Buchachenko:10}
\bibinfo{author}{\bibfnamefont{A.~A.} \bibnamefont{Buchachenko}},
  \bibinfo{author}{\bibfnamefont{G.}~\bibnamefont{{Cha{\l}asi\'nski}}},
  \bibnamefont{and} \bibinfo{author}{\bibfnamefont{M.~M.}
  \bibnamefont{{Szcz\c{e}\'sniak}}}, \bibinfo{journal}{J. Chem. Phys.}
  \textbf{\bibinfo{volume}{132}}, \bibinfo{pages}{024312}
  (\bibinfo{year}{2010}).

\bibitem[{\citenamefont{Cheeseman et~al.}(1990)\citenamefont{Cheeseman, {Van
  Zee}, Flanagan, and Weltner}}]{Cheeseman:90}
\bibinfo{author}{\bibfnamefont{M.}~\bibnamefont{Cheeseman}},
  \bibinfo{author}{\bibfnamefont{R.~J.} \bibnamefont{{Van Zee}}},
  \bibinfo{author}{\bibfnamefont{H.~L.} \bibnamefont{Flanagan}},
  \bibnamefont{and} \bibinfo{author}{\bibfnamefont{W.}~\bibnamefont{Weltner}},
  \bibinfo{journal}{J. Chem. Phys.} \textbf{\bibinfo{volume}{92}},
  \bibinfo{pages}{1553} (\bibinfo{year}{1990}).

\bibitem[{\citenamefont{Preuss}(2014)}]{Preuss:14}
\bibinfo{author}{\bibfnamefont{K.~E.} \bibnamefont{Preuss}},
  \bibinfo{journal}{Polyhedron} \textbf{\bibinfo{volume}{79}},
  \bibinfo{pages}{1} (\bibinfo{year}{2014}).

\bibitem[{\citenamefont{Szalay and Bartlett}(1993)}]{Szalay:93}
\bibinfo{author}{\bibfnamefont{P.~G.} \bibnamefont{Szalay}} \bibnamefont{and}
  \bibinfo{author}{\bibfnamefont{R.~J.} \bibnamefont{Bartlett}},
  \bibinfo{journal}{Chem. Phys. Lett.} \textbf{\bibinfo{volume}{214}},
  \bibinfo{pages}{481} (\bibinfo{year}{1993}).

\bibitem[{\citenamefont{Mou et~al.}(2017)\citenamefont{Mou, Tian, and
  Kertesz}}]{Mou:17}
\bibinfo{author}{\bibfnamefont{Z.}~\bibnamefont{Mou}},
  \bibinfo{author}{\bibfnamefont{Y.-H.} \bibnamefont{Tian}}, \bibnamefont{and}
  \bibinfo{author}{\bibfnamefont{M.}~\bibnamefont{Kertesz}},
  \bibinfo{journal}{Phys. Chem. Chem. Phys.} \textbf{\bibinfo{volume}{19}},
  \bibinfo{pages}{24761} (\bibinfo{year}{2017}).

\bibitem[{\citenamefont{Misquitta and Szalewicz}(2005)}]{Misquitta:05}
\bibinfo{author}{\bibfnamefont{A.~J.} \bibnamefont{Misquitta}}
  \bibnamefont{and}
  \bibinfo{author}{\bibfnamefont{K.}~\bibnamefont{Szalewicz}},
  \bibinfo{journal}{J. Chem. Phys.} \textbf{\bibinfo{volume}{122}},
  \bibinfo{pages}{214109} (\bibinfo{year}{2005}).

\end{thebibliography}
 \end{document}